\documentclass[sigconf,natbib=false]{acmart}

\AtBeginDocument{%
  \providecommand\BibTeX{{%
    \normalfont B\kern-0.5em{\scshape i\kern-0.25em b}\kern-0.8em\TeX}}}

\setcopyright{none}
\copyrightyear{2023}

\renewcommand\footnotetextcopyrightpermission[1]{} 
\acmISBN{}
\acmDOI{}

\input{misc/packages}
\newif\ifshowcomment
\showcommenttrue

\ifshowcomment
	\newcommand{\todo}[1]{\textsf{\color{red}{[{TODO: #1}]}}}
	\newcommand{\pratyush}[1]{\textsf{\color{orange}{[Pratyush: {#1}]}}}
    
\else
	\newcommand{\todo}[1]{}
	\newcommand{\pratyush}[1]{}
\fi

\def\sysRaw{Spork}
\def\sys{\sysRaw\xspace}
\def\sysE{SporkE\xspace}
\def\sysB{SporkB\xspace}
\def\sysC{SporkC\xspace}

\usepackage{siunitx}

\newcommand{\ms}{\si{\milli\second}\xspace}
\newcommand{\s}{\si{\second}\xspace}

\newcommand{\rev}[1] {#1}

\begin{document}

\title{Hybrid Computing for Interactive Datacenter Applications}

\author{Pratyush Patel}
\email{pratyush@cs.washington.edu}
\affiliation{%
  \institution{University of Washington}
  \country{USA}
}

\author{Katie Lim}
\email{katielim@cs.washington.edu}
\affiliation{%
  \institution{University of Washington}
  \country{USA}}

\author{Kushal Jhunjhunwalla}
\email{kushaljh@cs.washington.edu}
\affiliation{%
  \institution{University of Washington}
  \country{USA}
}

\author{Ashlie Martinez}
\email{ashmrtnz@cs.washington.edu}
\affiliation{%
 \institution{University of Washington}
 \country{USA}}

\author{Max Demoulin}
\email{maxdml@seas.upenn.edu}
\affiliation{%
  \institution{Astronomer}
  \country{USA}}

\author{Jacob Nelson}
\email{jacob.nelson@microsoft.com}
\affiliation{%
  \institution{Microsoft Research}
  \country{USA}}

\author{Irene Zhang}
\email{irene.zhang@microsoft.com}
\affiliation{%
  \institution{Microsoft Research}
  \country{USA}}

\author{Thomas Anderson}
\email{tom@cs.washington.edu}
\affiliation{%
  \institution{University of Washington}
  \country{USA}}

\renewcommand{\shortauthors}{Patel, et al.}


\begin{abstract}
Field-Programmable Gate Arrays (FPGAs) are more energy efficient and cost effective than CPUs for a wide variety of datacenter applications.
Yet, for latency-sensitive and bursty workloads, this advantage can be difficult to harness due to high FPGA spin-up costs.
We propose that a hybrid FPGA and CPU computing framework can harness the energy efficiency benefits of FPGAs for such workloads at reasonable cost.
Our key insight is to use FPGAs for stable-state workload and CPUs for short-term workload bursts. 
Using this insight, we design \sys, a lightweight hybrid scheduler that can realize these \rev{energy} efficiency and cost benefits in practice.
\rev{Depending on the desired objective, \sys can trade off energy efficiency for cost reduction and vice versa}.
It is parameterized with key differences between FPGAs and CPUs in terms of power draw, performance, cost, and spin-up latency. 
We vary this parameter space and analyze various application and worker configurations on production and synthetic traces.
Our evaluation of cloud workloads \rev{shows that energy-optimized \sys is not only more energy efficient but it is also cheaper than homogeneous platforms}---for short application requests \rev{with tight deadlines}, it is \rev{1.53}$\times$ more energy efficient and \rev{2.14}$\times$ cheaper than using only FPGAs. 
Relative to an idealized \rev{version of an existing} cost-optimized hybrid scheduler, energy-optimized \sys provides \rev{1.2--2.4}$\times$ higher energy efficiency at comparable cost, \rev{while cost-optimized \sys provides 1.1--2$\times$ higher energy efficiency at 1.06--1.2$\times$ lower cost.}
\end{abstract}




\settopmatter{printacmref=false, printccs=false, printfolios=false}
\maketitle

\section{Introduction}
\label{sec:introduction}



The rapidly growing energy needs of cloud applications have led to a renewed focus on energy efficiency by datacenter operators and their customers.
FPGAs use lower energy and provide better performance per dollar than
CPUs for many important datacenter
applications~\autocite{caulfield2016_Cloudscale,chung2018_Serving,istvan2017_Caribou,skhiri2019_FPGA,boutros2020_Peak,nurvitadhi2017_Can}.
However, offloading latency-sensitive and bursty applications onto a shared pool of FPGAs is complicated by their high spin-up costs.
To spin up new processing instances in response to application request bursts, FPGAs must be \emph{reconfigured}, a slow and energy-intensive process that can take several seconds~\autocite{intel_Stratix,caulfield2016_Cloudscale,du2022_Serverless}.
Such spin-up delays often violate request deadlines. 
Recurring spin-up costs can also outweigh the efficiency benefits.
Hence, FPGAs need to be overprovisioned to serve latency-sensitive and bursty traffic~\autocite{putnam2014_Reconfigurable,fowers2018_Configurable}.

Our goal in this paper is to show how to use a hybrid FPGA and CPU platform to harness the energy efficiency benefits of FPGAs for latency-sensitive and bursty applications even at reasonable cost.
We observe that FPGAs and CPUs offer \emph{complementary benefits}: FPGAs typically use less energy and perform better for steady workloads whereas CPUs can adapt to workload variations and are cheaper at low load. 
A hybrid platform that interchangeably runs application requests can be more efficient and cheaper than using FPGAs or CPUs alone.

Hybrid platforms face two key challenges. 
First, programmers must develop and maintain application implementations for both FPGA and CPU workers~\autocite{intel_OneAPI,zhang2022_HeteroGen,chen2018_TVM,tine_SingleSource}.
Second, orchestration frameworks need to be modified to run requests across heterogeneous workers~\autocite{du2022_Serverless,inaccel_FPGA,tarafdar2019_LibGalapagos,ringlein2020_Programming,ringlein2019_System}.
Although some prior work attempts to address these issues, additional effort is necessary to enable hybrid FPGA-CPU computing in practice.
Therefore, we analyze whether the energy efficiency and cost benefits offered by hybrid platforms could justify these efforts.

%

We propose a hybrid FPGA-CPU computing framework that uses FPGAs to serve stable-state workload and CPUs for request bursts that occur at a higher frequency than the FPGA spin-up time.
Under certain assumptions about the scheduler, workload, and computational workers, we show that for applications with moderate-to-high burstiness, such a framework can offer \rev{higher} energy efficiency \rev{than} FPGA-only platforms while being much cheaper.
\rev{Appropriately configured, a hybrid platform can further trade off these efficiency gains for cost reduction as desired.}

We design \sys, a \rev{lightweight} hybrid computing scheduler that realizes these benefits in practice, given an adequate supply of FPGA and CPU workers. 
\sys proactively allocates FPGA workers in each scheduling interval, based on the historical distribution of the number of workers required to service the load. 
\rev{Allocations are chosen such that they minimize a weighted sum of the expected energy usage and cost in the subsequent interval.}
\sys then tunes allocations within the interval to ensure that request deadlines are met, assuming request sizes are known in advance. 
Specifically, it handles workload bursts and FPGA under-allocations (due to misprediction) by dynamically allocating short-lived CPU workers on the request dispatch path.
Finally, \sys dispatches requests such that FPGA utilization is maximized and idle workers are reclaimed quickly to reduce energy usage and costs.

Since different FPGA and CPU workers have different characteristics that are also affected by application behaviour,
we parameterize \sys with the key differences in worker power draw, performance, costs, and spin-up overheads.
By varying this parameter space, we perform the first detailed study of the energy efficiency and cost trade-offs between hybrid and homogeneous platforms across a large set of worker and workload configurations on production and synthetic traces.

Our evaluation of cloud workloads shows that \sys is significantly more energy efficient and cheaper than homogeneous platforms, given plausible assumptions about deadlines and the relative FPGA-to-CPU performance and energy ratios.
For production traces with \rev{tight deadlines}, \rev{energy-optimized} \sys is \rev{1.53$\times$} more energy efficient and costs \rev{2.14$\times$} lower than using only FPGAs. 
Compared to an idealized version of MArk~\autocite{zhang2019_MArk}, a state-of-the-art cost-optimized hybrid scheduler, \rev{energy-optimized} \sys provides \rev{1.2--2.4}$\times$ higher energy efficiency at similar cost \rev{whereas cost-optimized \sys provides 1.1--2$\times$ higher energy efficiency at 1.06--2$\times$ lower cost.}
\rev{Energy-optimized} \sys uses only \rev{7--14}\% more energy and \rev{cost-optimized \sys incurs only 5--17\% higher cost} than an idealized, best case FPGA-only system with no spin up and idling overheads.


\smallskip
\noindent We make the following contributions:
\begin{itemize}[nosep]
	\item We motivate hybrid FPGA-CPU computing by showing that under idealized settings, hybrid platforms can harness the energy efficiency benefits of FPGAs at a lower cost for latency-sensitive and bursty workloads~(\Cref{sec:motivation}). 
    \item \rev{By configuring the optimization objective, we show that hybrid platforms can incrementally trade off these energy efficiency gains for higher cost reduction~(\Cref{sec:motivation}).}
	\item We design \sys, \rev{a practical hybrid scheduler that realizes these benefits in practice using lightweight prediction and efficient request dispatch}~(\Cref{sec:orchestration}).
	\item \rev{We evaluate three variants of \sys on production and synthetic traces and show that they} substantially outperform state-of-the-art homogeneous and hybrid platforms under several plausible workload and worker configurations~(\Cref{sec:evaluation}).
\end{itemize}

\section{Background}
\label{sec:background}


\subsection{Datacenter Platforms}

\begin{figure}
	\centering
	\resizebox{0.75\columnwidth}{!}{
		\includegraphics{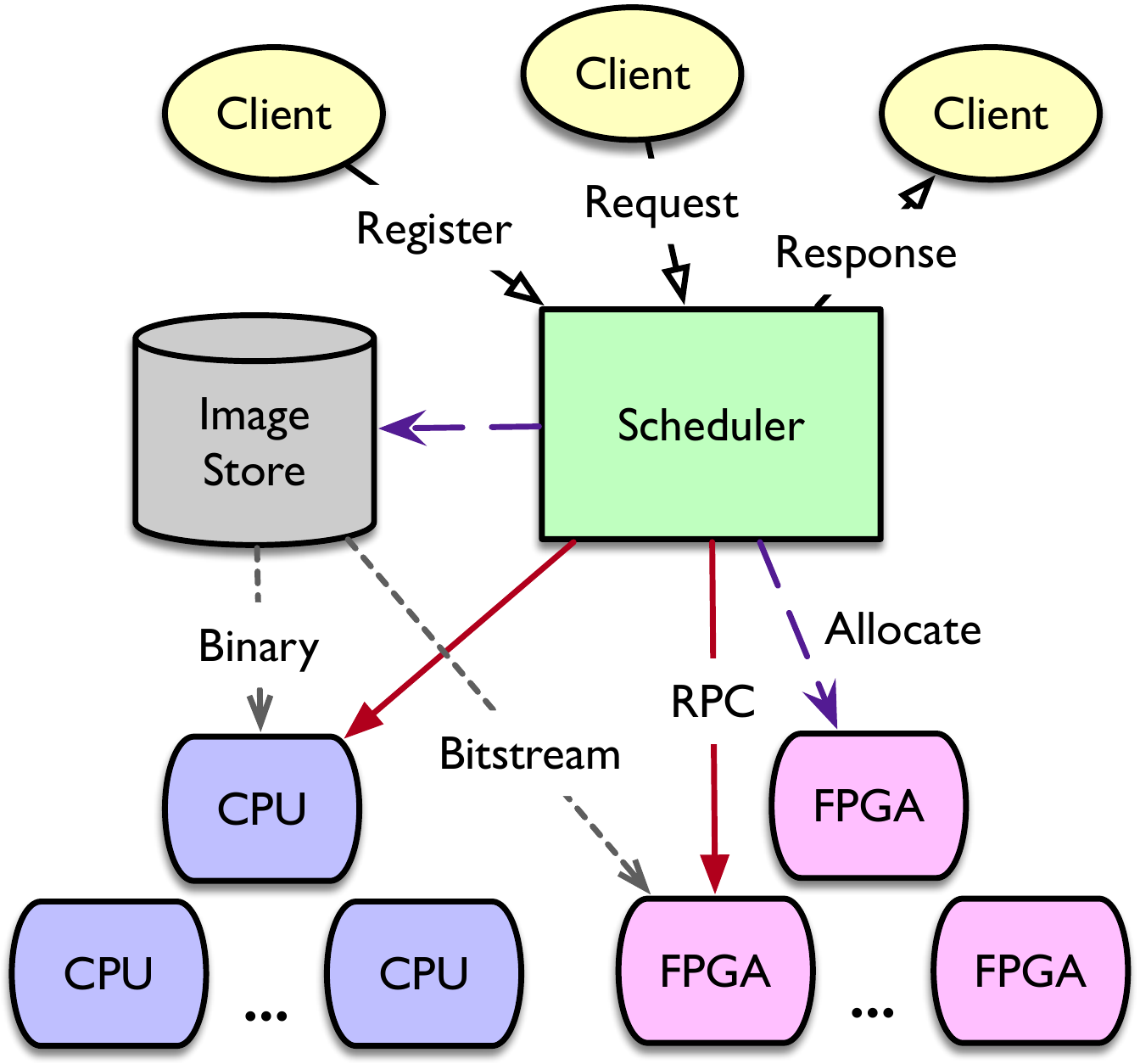}
	}
	\caption{\sys Architecture.}
	\label{fig:design_overview}
\end{figure}


\Cref{fig:design_overview} shows the high-level architecture of \sys, similar in design to serverless frameworks~\autocite{openfaas_OpenFaaS,apache_OpenWhisk,jia2021_Nightcore,du2022_Serverless}.
We target latency-sensitive and bursty stateless functions running on a shared pool of FPGA and CPU workers~\autocite{martin_Microservices,shahrad2020_Serverless,jonas2019_Cloud,wang2018_Peeking,CNCF_serverless}.
Clients invoke applications by sending requests to a scheduler, which performs two key operations:
(1)~it allocates (spins up) and deallocates (spins down) workers from the pool to handle incoming load, and
(2)~it dispatches requests to allocated workers and returns results back to clients.
Application requests must adhere to pre-specified latency deadlines---for example, clients may require that requests complete within 10$\times$ the request service time~\autocite{mogul2019_Nines}.

\subsection{FPGA and CPU Trade-offs}
\label{sec:cpus_vs_fpgas}

FPGAs and CPUs differ in several dimensions that can impact the energy efficiency and cost of the platform. 
We discuss these below.

\begin{table}
	\footnotesize
	\centering
	\begin{tabular}{@{}cc|cc@{}}
		\toprule
		\multicolumn{2}{c|}{\textbf{CPU}}         & \multicolumn{2}{c}{\textbf{FPGA}}                                                                      \\
		Deployment                                & Spin up                            & Reconfiguration                                  & Spin up          \\ \midrule
		Container~\autocite{manco2017_My}         & $\sim$500ms                       & Flash                                            & $\sim$1--3min   \\
		MicroVM~\autocite{agache2020_Firecracker} & $\sim$125ms                       & JTAG~\autocite{intel_Stratix}                    & $\sim$10--30s   \\
		Process~\autocite{du2020_Catalyzer}       & $\sim$5ms                         & PCIe~\autocite{amazonwebservices_EC2}            & $\sim$1--10s    \\
		Co-routine~\autocite{cloudflare_Workers}  & \textless 1ms                     & ICAP*~\autocite{korolija2020_OS,mishra2016_REoN} & $\sim$500ms \\ \bottomrule
	\end{tabular}
	\caption{Spin-up latencies for different CPU and FPGA deployments.
		*ICAP is a partial reconfiguration technique that does not work with arbitrary application logic.}
	\label{tab:device_spinups}
\end{table}

\begin{table}
	\footnotesize
	\centering
	\begin{tabular}{@{}ccc@{}}
		\toprule
		\textbf{Application}
		                                                   &
		\textbf{Power Efficiency}                           &
		\textbf{Speedup}
		\\
		\midrule
		CNN Inference~\autocite{sharma2016_Highlevel}
		                                                   & 2.1--5.3$\times$
		                                                   &
		1--7.7$\times$
		\\
		RNN Inference~\autocite{nurvitadhi2016_Accelerating}
		                                                   & 1.8--4.5$\times$
		                                                   &
		0.75--4$\times$
		\\
		Web Search~\autocite{putnam2014_Reconfigurable}
		                                                   & N/A
		                                                   & 1.95$\times$
		\\
		Image Processing~\autocite{qasaimeh2019_Comparing} &
		5.68$\times$
		                                                   & N/A
		\\
		Intrusion Detection~\autocite{zhao2020_Achieving}
		                                                   & 1.4--3$\times$
		                                                   &
		5.8--8.8$\times$
		\\
		Document Filtering~\autocite{chen2012_Using}
		                                                   & 6.19$\times$
		                                                   & 0.85$\times$
		\\
		Video Compression~\autocite{chen2013_Fractal}
		                                                   & 5.2$\times$
		                                                   & 8.4--10.1$\times$
		\\
		Bzip2 Compression~\autocite{qiao2019_FPGABased}
		                                                   & N/A
		                                                   & 1.6--2.3$\times$
 	\\
		\rev{HPC Kernels~\autocite{nguyen2022fpga}}
		                                                   & \rev{2--6$\times$}
		                                                   & \rev{0.07--1$\times$}  
		\\
		Key-value Stores~\autocite{lavasani2014_FPGAbased}
		                                                   & 2.18$\times$
		                                                   & 1.02$\times$
		\\
		Databases~\autocite{istvan2017_Caribou}
		                                                   & 1.42$\times$
		                                                   & 1--2$\times$
		\\ \midrule
		\rev{Geometric Mean}                                   & \rev{3.32$\times$}      & \rev{1.88$\times$}
		\\ \bottomrule
	\end{tabular}
	\caption{FPGA busy power efficiency and processing speedups over CPUs for various applications. Values are derived from cited works; most compare single-socket CPUs with equivalent FPGAs.}
	\label{tab:app}
    \vspace{-\baselineskip}
\end{table}

\medskip
\noindent\textbf{Spin-up Latency.} FPGA workers take longer and consume more energy to spin up than CPUs~\autocite{du2020_Catalyzer,cloudflare_Eliminating,intel_Stratix,du2022_Serverless}.
In practice, spin-up latency depends on the deployment model as shown
in~\Cref{tab:device_spinups}.
Reconfiguration is the key latency bottleneck for FPGAs.
Frequent FPGA spin ups also increase energy usage since reconfiguration is power intensive and no useful work can be done in parallel.\footnote{Existing logic
on FPGAs can be used while a new application bitstream is being written to Flash
memory.} 
Partial reconfiguration is quicker and can complete under a second~\autocite{mishra2016_REoN,korolija2020_OS}. However, due to FPGA design and floor planning restrictions, it is not usable with arbitrary application logic~\autocite{intel_Stratixa}, and so we do not consider it here.
These factors lead to FPGAs being statically provisioned for most datacenter applications~\autocite{putnam2014_Reconfigurable,amazonwebservices_EC2}, whereas CPUs are often dynamically provisioned to match load~\autocite{shahrad2020_Serverless,jia2021_Nightcore,cloudflare_Eliminating,fuerst2021_FaasCache}.

\medskip
\noindent\textbf{Efficiency and Performance.} FPGAs typically consume less energy than CPUs to run the same applications. Energy depends on the power and execution duration. FPGAs tend to perform better at both due to specialized implementations. Specialization eliminates inefficient instruction fetch and decode pipelines present in CPUs and lets FPGAs exploit parallelism best suited to each application.
\Cref{tab:app} lists a number of studies that show a benefit from FPGA offload. Note that these are run on different hardware generations. For these applications and hardware configurations, FPGAs are, on average, \rev{$\sim$6.25$\times$ (3.32 $\times$ 1.88)} more energy efficient than equivalent single-socket CPUs. Even when an application has lower performance on an FPGA (e.g., 
document filtering), it tends to be more energy efficient overall.

\medskip
\noindent\textbf{Idle Power.} Spun-up workers draw power even when idle. 
Energy proportionality, the ratio between idle and active power usage, is often much better for CPUs. For example, idle CPUs today use only $\sim$15\% of the power drawn at maximum utilization~\autocite{barroso2018_Datacenter}.
Most FPGAs do not inherently support power reduction techniques such as Dynamic Voltage and Frequency Scaling (DVFS) and sleep states, although specific designs can be adopted to reduce idle power waste~\autocite{hosseinabady2014_Runtime,zhao2016_Universal}.
However, since CPUs have higher maximum wattages, both platforms tend to have comparable idle power draw.



\medskip
\noindent\textbf{Costs.} FPGAs usually incur higher capital expenses than comparable CPUs, and they are more expensive per worker instance.
For example, on AWS cloud~\autocite{AWS_EC2_pricing}: an 8 vCPU virtual machine (x2gd.2xlarge) costs 67\textcent/hr, whereas an F1 FPGA (f1.2xlarge minus equivalent CPU host VM) costs 98\textcent/hr ($\sim$1.5$\times$ difference). Similar trends also hold for Alibaba Cloud's F3 FPGA instances~\autocite{alibaba_Deep,alibaba_pricing}. 



\section{The Case for Hybrid Computing}
\label{sec:motivation}

We develop insight into hybrid computing by evaluating latency-sensitive and bursty workloads on hybrid and homogeneous platforms in an idealized setting where task arrivals are known in advance.
In the following section, we use these insights to develop a practical and efficient hybrid scheduler without this assumption.

\subsection{Pareto-Optimal Schedulers}

\Cref{tab:energy_optimal_oracle} shows the parameters for an energy-optimal hybrid scheduler that uses perfect workload information to minimize the overall energy used by an application. 
The scheduler solves a Mixed-Integer Linear Program (MILP) to determine the number and type of workers to allocate over time. 
The MILP constraints ensure that: 
(1)~sufficient workers are allocated in each scheduling interval to satisfy all requests arriving in that interval, and
(2)~the most efficient decision between deallocating workers or keeping them idle is made, given knowledge of future load.
We reuse the same formulation for homogeneous platforms, but constrain allocatable worker types.
We include an additional constraint for FPGAs to ensure they are allocated for least as long as the spin-up duration.
\rev{The cost-optimal scheduler has a similar formulation; rather than differentiating between idle and busy workers, it only considers the duration for which workers are spun up to estimate cost.
Other pareto-optimal schedulers minimize weighted sums of energy and cost.}

Beyond having perfect future workload information, these schedulers make two impractical assumptions. 
First, they assume that worker allocations and deallocations are instantaneous; they still incur energy and cost overheads. 
Second, they assume that all requests that arrive also finish within the same interval, with no spillover across intervals. 
\rev{Together, these assumptions simplify constraint formulation and help provide a rough bound for hybrid computing's efficiency and cost benefits over homogeneous platforms.}
In particular, they are optimized for FPGAs---in practice, some misprediction is likely, leading FPGA-only platforms to require overprovisioning, delay tolerance, or both.


\begin{table}
	\footnotesize
	\centering
	\begin{tabular}{cl}
		\toprule
		\multicolumn{2}{c}{\textbf{Inputs}}                                                                                           \\
		$X_{t} \in [0,\infty)$                       & number of requests that arrive in time $[t, t+1)$                              \\
		$a^{c}, a^{f}$                               & allocation energy of CPU, FPGA                                                 \\
		$d^{c}, d^{f}$                               & deallocation energy of CPU, FPGA                                               \\
		$e^{c}_{b}, e^{f}_b$                         & busy power of CPU, FPGA within an interval                                     \\
		$e^{c}_{i}, e^{f}_i$                         & idle power of CPU, FPGA within an interval                                     \\
		$r^{c}, r^{f}$                               & request processing rates of CPU, FPGA                                         \\ 
		$S$                                          & spin-up latency of FPGA                                                      \\ 
  \midrule
		\multicolumn{2}{c}{\textbf{Outputs}}                                                                                          \\
		$Y^{c}_{t} \in [0, N_{c}]$                   & total CPUs allocated in time $[t,t+1)$                                         \\
		$Y^{f}_{t} \in [0, N_{f}]$                   & total FPGAs allocated in time $[t,t+1)$                                        \\
		$B^{c}_{t}, B^{f}_{t} \in [0, N_{r}]$        & busy CPUs, FPGAs in time $[t,t+1)$                                             \\ \midrule
		\multicolumn{2}{c}{\textbf{Helpers}}                                                                                          \\
		$I^{c}_{t}, I^{f}_{t} \in [0, N_{r}]$        & idle CPUs, FPGAs in time $[t,t+1)$                                             \\
		$E^{a}_{t}, E^{d}_{t}, E^{b}_{t}, E^{i}_{t}$ & total alloc, dealloc, busy, idle energy in $[t,t+1)$                           \\ \midrule
		\multicolumn{2}{c}{\textbf{Constraints}}                                                                                      \\
		$\min$                                       & $\sum_{t}(E^{a}_{t} + E^{b}_{t} + E^{i}_{t} + E^{d}_{t})$                      \\
		s.t.                                         & $\forall t: E^{a}_{t} = \sum_{w\in\{c, f\}}\max(Y^{w}_{t+1} - Y^{w}_{t}, 0) a^w$ \\
		                                             & $\forall t: E^{b}_{t} = \sum_{w\in\{c, f\}} e^{w}_{b}B^{w}_{t}$                  \\
		                                             & $\forall t: E^{i}_{t} = \sum_{w\in\{c, f\}} e^{w}_{i}I^{w}_{t}$                  \\
		                                             & $\forall t: E^{d}_{t} = \sum_{w\in\{c, f\}}\max(Y^{w}_{t} - Y^{w}_{t+1}, 0) d^w$ \\
		                                             & $\forall t: r^{c}B^{c}_{t} + r^{f}B^{f}_{t} = X_{t}$                        \\
		                                             & $\forall t: Y^{c}_{t} = B^{c}_{t} + I^{c}_{t} \le N_{c}$                       \\
		                                             & $\forall t: Y^{f}_{t} = B^{f}_{t} + I^{f}_{t} \le N_{f}$                       \\
		                                             & $\forall t: Y^{f}_{t+S} \geq \sum_{\tau=t}^{t+S} \max(Y^{f}_{\tau+1} - Y^{f}_{\tau}, 0) $                 \\
		\bottomrule
	\end{tabular}
	\caption{\rev{Parameters for a Mixed-Integer Linear Program (MILP) scheduler that minimizes energy usage
		across CPU ($c$) and FPGA ($f$) workers in each epoch with perfect workload knowledge.
		Cost-optimal formulation is similar; other pareto-optimal formulations minimize weighted sums of energy and cost.}}
	\label{tab:energy_optimal_oracle}
    \vspace{-\baselineskip}
\end{table}


\subsection{Hybrid vs. Homogeneous Platforms}

\medskip
\noindent\textbf{Assumptions.}
Based on our survey in~\Cref{sec:cpus_vs_fpgas}, we assume that FPGA workers are \rev{6$\times$} more energy efficient than CPUs when running requests (\rev{150W CPU}, 50W FPGA, FPGAs are \rev{2$\times$} faster) and draw lower power when idle (30W CPU, 20W FPGA~\autocite{nguyen2022fpga}). 
CPU workers take \rev{0.75} J to spin up (5\ms to spin up at \rev{150W}) whereas FPGA workers take 500 J for reconfiguration (10\s to spin up at 50W).
Workers are priced such that FPGAs are 1.5$\times$ more expensive to use than CPUs (67\textcent/hr CPU, 98\textcent/hr FPGA). Note that this cost is different than the total cost of ownership (TCO)~\autocite{barroso2018_Datacenter}---it simply represents the worker occupancy cost for a fixed time duration.

We evaluate self-similar workload traces because such distributions are commonly observed in the cloud~\autocite{yin2015_BURSE,leland1994_Selfsimilar,pitchumani2015_Realistic,tootaghaj2015_Evaluating}. Specifically, we generate per-second request arrival rates using the b-model~\autocite{wang2002_Data}, where the bias parameter captures workload burstiness.
For intuition, a burstiness value of 0.5 corresponds to a uniform trace, whereas 0.75 implies very high variability (over $\sim$20$\times$ difference in load for some consecutive intervals).
Each trace is an hour long and has an average rate of 10,000 requests per second.
Request sizes are constant and run within 10\ms on CPU workers. All requests finish in the same interval that they arrive.
Workers are abundant and scale up and down to match incoming demand.


\begin{figure}
	\centering
	\subfloat[Energy-optimal scheduling.]{
		\includegraphics[width=3.375in]{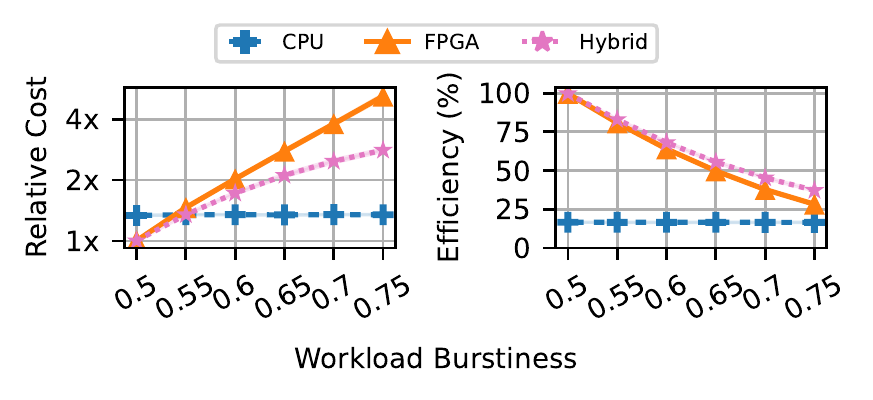}}\\
	\subfloat[Cost-optimal scheduling.]{
		\includegraphics[width=3.375in]{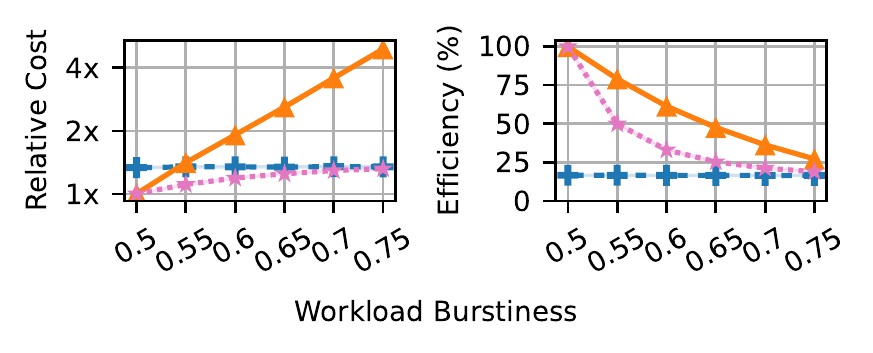}}
	\caption{\rev{Energy efficiency and cost of CPU-only, FPGA-only, and hybrid
		platforms with increasing workload burstiness using an optimal, rate-based scheduler}.
        Results are normalized to an idealized FPGA-only platform with no overhead, averaged over ten trace runs. Self-similar traces are generated using the b-model~\autocite{wang2002_Data}, where b=0.5 represents uniform load and b=0.75 is highly variable.}
	\label{fig:motivation_cost_efficiency}
    \vspace{-1.3\baselineskip}
\end{figure}

\begin{figure}
	\centering
	\includegraphics[width=3in]{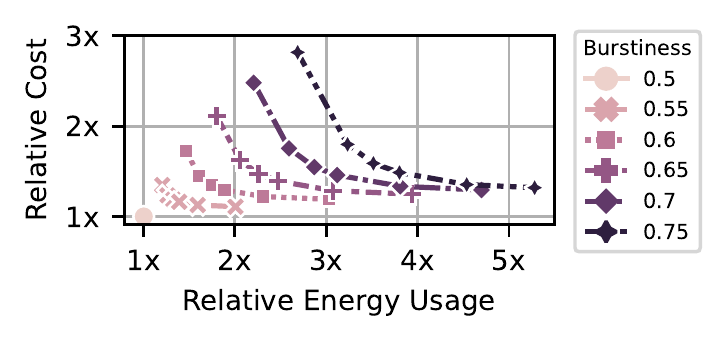}
	\caption{\rev{Energy usage and cost trade-offs among pareto-optimal MILP schedulers at different workload burstiness, relative to an idealized FPGA-only platform. Boundary markers for each burstiness value represent energy-optimal and cost-optimal schedulers.}}
	\label{fig:motivation_pareto}
    \vspace{-\baselineskip}
\end{figure}

\medskip
\noindent\textbf{Results.}
\Cref{fig:motivation_cost_efficiency} shows the energy efficiency and cost of homogeneous and hybrid platforms with increasing workload burstiness.
Results are relative to an idealized FPGA-only platform that only incurs compute costs (i.e., zero spin up and idling overheads). Each data point is averaged across ten randomly generated trace runs.

Among homogeneous platforms, we find that FPGAs are better at serving
workloads with low burstiness (up to \rev{1.3$\times$ cheaper and 6$\times$} more energy
efficient than CPUs), whereas CPUs work well at high burstiness (up to
\rev{4$\times$ cheaper} and only 1.7$\times$ less energy efficient than FPGAs). 
In contrast, hybrid platforms get the best of both worlds---better than or equivalent to homogeneous platforms for the optimization metric.

\rev{In an energy-optimal setting, FPGA-only and hybrid provide similar energy efficiency at low burstiness since it is desirable to serve most requests using FPGAs. 
Yet, hybrid is up to \rev{1.3$\times$} more energy efficient and up to 1.8$\times$ cheaper at high burstiness since it can also use CPUs to handle transient bursts. 
At low-to-moderate burstiness, hybrid dominates---it provides slightly higher energy efficiency and costs lower than CPU or FPGA-only.}

\rev{Cost-optimized hybrid platforms are less expensive than either homogeneous platform while being much more energy efficient than CPU-only. 
Compared to using only FPGAs, hybrid platforms at high workload burstiness are up to 3.8$\times$ cheaper while being only 1.4$\times$ less energy efficient. 
This advantage declines at moderate burstiness where the hybrid approach is up to 1.6$\times$ cheaper and 1.8$\times$ less energy efficient.}


Homogeneous platforms produce similar results whether energy or cost optimized. 
In contrast, hybrid platforms can trade off one metric for the other \rev{by weighting the desired objective, as shown by the pareto-optimal curves in~\Cref{fig:motivation_pareto}.} 
This trade-off can be significant. 
For example, energy-optimal hybrid platforms at high burstiness are \rev{over} 2$\times$ more expensive than cost optimal.
Because FPGAs are much more energy efficient than CPUs, the energy-optimal solution is to spin up an FPGA even at low average utilization, incurring extra cost.
In a cost-optimal \rev{setting}, FPGAs will only be used if the workload is heavy enough to make it cost efficient to do so.
\section{Practical Hybrid Scheduling}
\label{sec:orchestration}

To make hybrid FPGA-CPU computing practical, we need:
(1)~an orchestration framework that manages heterogeneous workers and assigns application requests, and
(2)~a scheduler that makes worker allocation and request dispatching decisions to achieve desired energy efficiency and cost trade-offs.
Prior work has developed frameworks to support FPGA and CPU worker management~\autocite{du2022_Serverless,inaccel_FPGA,tarafdar2019_Building} and interchangeable application execution in specific settings~\autocite{ringlein2020_Programming,tarafdar2019_LibGalapagos}.
Although not complete hybrid orchestration platforms, these efforts provide the building blocks to enable hybrid FPGA-CPU computing.
Our focus is on the second piece---\rev{to design a practical hybrid scheduler that can trade off energy efficiency and cost and determine configurations under which hybrid computing provides substantial benefits.}

\begin{table}
	\footnotesize
	\centering
	\begin{tabular}{@{}ccccc@{}}
		\toprule
		\textbf{Scheduler}                           & \textbf{Hybrid?} & \textbf{Pred?} & \textbf{Alloc Freq.} & \textbf{Dispatch} \\ \midrule
		AutoScale~\autocite{gandhi2012_AutoScale}    & No                 & No             & Long-term            & Index packing \\
		Serverless~\autocite{shahrad2020_Serverless} & No                 & No             & Short-term           & 1:1               \\
		INFaaS~\autocite{romero2019_INFaaS}          & Yes                & No             & Long-term            & Utilization-based \\
		MArk~\autocite{zhang2019_MArk}               & Yes                & Costly         & Both                 & Round robin       \\
		\sys                                         & Yes                & Cheap          & Both                 & Efficient-first   \\ \bottomrule
	\end{tabular}
	\caption{Summary of schedulers.}
	\label{tab:existing_schedulers}
    \vspace{-\baselineskip}
\end{table}

We first consider whether existing schedulers can address this challenge.~\Cref{tab:existing_schedulers} summarizes design choices made by existing schedulers.
AutoScale~\autocite{gandhi2012_AutoScale}
and Serverless~\autocite{shahrad2020_Serverless} are homogeneous schedulers
which \rev{reduce} energy use and strike a balance between worker spin ups and idling.
AutoScale also introduces energy efficient index-based dispatching, where workers are packed up to a threshold in busiest-first order to make it easier to spin down idle workers. 
INFaaS~\autocite{romero2019_INFaaS} and MArk~\autocite{zhang2019_MArk} provide
cost-optimized, \rev{deadline-aware} hybrid schedulers for inference serving on CPUs and GPUs. 
INFaaS relies on reactive heuristics that leave efficiency gains on the table. MArk uses LSTM-based prediction which can be costly and difficult to scale.
MArk also uses inefficient round robin dispatch (see~\Cref{sec:scheduling_efficacy}).
Crucially, \rev{neither considers energy efficiency}, nor do they account for differences between idle and busy workers. 
We bridge this gap by proposing \sys, a lightweight hybrid scheduler that models these worker properties and \rev{optimizes both the energy consumption and cost while meeting request deadlines.}

\subsection{\sys Overview}

\rev{For simplicity of exposition, we describe \sys primarily from an energy optimization perspective and later discuss extensions to incorporate cost optimization and trade offs.}

\sys performs two key functions: worker allocation/deallocation and request dispatch. 
\sys's underlying strategy is to use FPGAs for steady-state requests and CPUs for short-term bursts and low load.
It attempts to right-size FPGA allocations per scheduling interval by predicting the number of workers needed and allocating them at the start of the interval (lower bounded by the FPGA spin-up latency). 
\rev{\sys's predictor makes estimations using:
(1)~the historical distribution of required FPGA worker counts conditioned on worker counts required in recent intervals, and
(2)~the amortized worker spin-up overhead over its expected lifetime conditioned on currently allocated worker count.}
Specifically, it allocates the worker count value that minimizes the expected sum of idle and busy energy consumption across all candidates in the conditional distribution.
If needed, \sys compensates for any FPGA under-allocations or request bursts by allocating short-lived CPU workers on the request dispatch path, like in typical serverless orchestration~\autocite{openfaas_OpenFaaS,apache_OpenWhisk,jia2021_Nightcore}.
Workers are deallocated after keeping them idle for a fixed duration as an insurance against inefficient, repetitive allocations~\cite{gandhi2012_AutoScale,shahrad2020_Serverless}.



For request dispatch, our strategy is to schedule requests such that (1)~they meet their deadlines, (2)~FPGA worker utilization is maximized, and (3)~idle workers are reclaimed quickly. These criteria help meet latency requirements and reduce energy usage and costs. To meet deadlines, our dispatcher estimates request sizes and worker completion times before making dispatch decisions. To maximize energy efficiency, it dispatches requests to FPGAs over less efficient CPUs. To quickly reclaim idle workers, it packs as much load as possible on the busiest workers, which frees up other workers to be reclaimed after the idle timeout~\autocite{gandhi2012_AutoScale}.


\begin{table}
	\footnotesize
	\centering
	\begin{tabular}{@{}ll@{}}
		\toprule
		\textbf{Symbol} & \multicolumn{1}{c}{\textbf{Description}} \\ \midrule
		$A_{w}$         & Spin-up latency for worker $w$     \\
		$E_{w}$         & Average request service time on worker $w$     \\
		$B_{w}$         & Runtime busy power for worker $w$        \\
		$I_{w}$         & Idle power for worker $w$                \\
		\rev{$C_{w}$}    & \rev{Cost per time unit for worker $w$}            \\
		$S$             & Speedup factor of FPGA over CPU          \\
		$n_{t}$         & Number of FPGAs required in interval $t$ \\
		$T_{s}$         & Scheduling interval length \\
		$T_{b}$         & Breakeven service threshold for FPGA efficiency over CPU \\
		\bottomrule
	\end{tabular}
	\caption{Summary of notations. Subscript $w \in \{c, f\}$ for CPU and FPGA workers respectively.}
	\label{tab:notations}
    \vspace{-\baselineskip}
\end{table}

\Cref{tab:notations} lists notations used in this section.

\subsection{Per-Interval Allocation}
\label{sec:fpga_allocation}


\Cref{alg:per_interval_alloc} describes \sys's per-interval allocator. 
\sys adjusts FPGA worker allocations periodically due to their long spin-up times. 
Allocation periods ($T_{s}$) are lower bounded by the FPGA spin-up time ($A_{f}$). 
For simplicity, we assume that $T_{s} = A_{f}$ in our description.
At the start of any interval $t$, \sys uses the number of FPGA workers needed in the previous interval, $n_{t-1}$, \rev{the historical distribution of worker counts, $\mathbb{H}$, and the historical distribution of worker lifetimes, $\mathbb{L}$,} to predict and allocate the number of FPGAs needed in the subsequent interval, $n_{t+1}$. 
\sys predicts $n_{t+1}$ rather than $n_{t}$ to have the time to spin up any needed FPGAs.

\begin{algorithm}
	\footnotesize
	\caption{\rev{Per-interval FPGA allocation algorithm.}}
	\label{alg:per_interval_alloc}
	\begin{algorithmic}[1]
		\Procedure{Allocator}{}
		\State{$n_{t-1}$: \# of FPGA workers needed in the previous interval}
		\State{$n_{t+1}$: predicted \# of FPGA workers needed in the next interval}
        \State{$\mathbb{H} \gets$ hashmap of histograms that track the distribution of \# of FPGA workers needed in an interval, conditional on the value two intervals ago}
        \State{$\mathbb{L} \gets$ hashmap of the average lifetime of an FPGA worker, conditional on the number of workers already allocated}
		\State{$\mathcal{F}_{t-1} \gets$ sum of request service times on FPGAs in the previous interval}
		\State{$\mathcal{C}_{t-1} \gets$ sum of request service times on CPUs in the previous interval}
		\State{$n_{t-1} \gets$ \Call{NeededFPGAs}{$\mathcal{F}_{t-1}, \mathcal{C}_{t-1}$}} 
		\State{$\mathbb{H}[n_{t-3}].add(n_{t-1})$} \Comment{Update histogram}
		\State{$n_{t+1} \gets \Call{PredictFPGAs}{\mathbb{H}, \mathbb{L}, n_{t-1}}$} \label{alg:call_predict_num_fpgas}
		\State{\Call{AllocFPGAs}{$n_{t+1}$}}
		\State{\Call{Sleep}{$T_{s}$} and repeat} \Comment{$T_{s} = A_{f}$}
		\EndProcedure
  	\Function{NeededFPGAs}{$\mathcal{F}, \mathcal{C}$} \label{alg:static_opt:update_hist}
		\State{$\lambda \gets \mathcal{F} + \mathcal{C} / S  $} \Comment{Weighted sum of request sizes}
		\State{$n \gets \lfloor \frac{\lambda}{T_{s}} \rfloor $} 
		\If{$\lambda \bmod T_{s} > T_b$} \Comment{Breakeven threshold from~\Cref{eq:breakeven_threshold}}
		\State{$n \gets n + 1$}
		\EndIf
		\State{\textbf{return} $n$} 
		\EndFunction
	\end{algorithmic}
\end{algorithm}

\sys calculates $n_{t-1}$ by estimating the number of FPGA workers that would have been needed to optimally serve the aggregate request demand in that interval. 
To estimate aggregate request demand, \sys sums the request sizes on FPGA and CPU workers in the previous interval, weighted by the FPGA speedup factor, $S$.
It then divides the total by the interval length to calculate $n_{t-1}$.
\sys rounds up this estimate based on the breakeven threshold, $T_{b}$, the service time threshold beyond which running requests on an FPGA is more energy efficient than using a CPU. 
$T_{b}$ is given by:
\begin{equation}
	\rev{T_{b}B_{c} = \frac{T_{b}}{S}B_{f} + \left( T_{s} - \frac{T_{b}}{S} \right) I_{f},}
	\label{eq:breakeven_threshold}
\end{equation}
\rev{where, $B_{f}$, $B_{c}$, and $I_{f}$ are the FPGA and CPU busy and idle power draws, respectively. We assume that the idle energy overhead from CPU workers is negligible since they are only allocated briefly to serve request bursts.}

\sys then uses $n_{t-1}$ to predict the number of FPGAs needed in the subsequent interval, $n_{t+1}$, by invoking~\Cref{alg:predict_num_fpgas}.
The predictor uses \rev{$\mathbb{H}$, a hashmap} of histograms of historical worker counts, conditional on the worker count needed to service load two intervals prior (since FPGA allocation takes one interval). 
Map entries are indexed by the number of FPGA workers needed in an interval; each entry is a histogram that captures the distribution of worker counts needed two intervals later.
Histograms are binned by the number of required FPGA workers per interval and track the \rev{probability} with which they have been observed so far.
Every interval, the allocator updates (or initializes) the histogram at $\mathbb{H}[n_{t-3}]$ with the worker count needed in the previous interval, $n_{t-1}$. 
\rev{To account for spin-up overheads, the predictor also uses $\mathbb{L}$, a hashmap of the average lifetime of an FPGA worker, conditional on the number of workers already allocated. 
$\mathbb{L}$ is updated upon worker deallocations.}

%


\begin{algorithm}
	\footnotesize
	\caption{\rev{Estimating the most energy efficient FPGA allocation given the historical distribution of required worker counts.}}
	\label{alg:predict_num_fpgas}
	\begin{algorithmic}[1]
		\Procedure{PredictFPGAs}{$\mathbb{H}, \mathbb{L}, n_{t-1}$}
		\State{$n_{curr}$: currently allocated \# of FPGAs}
		\State{$n_{t+1}$: predicted \# of FPGA workers for the next interval}
  	\If{$n_{t-1} \notin \mathbb{H} $} \Comment{If we have not seen this count before}
		\State{$n_{t+1} \gets n_{t-1}$} \Comment{Maintain previously needed worker count}
		\State{\textbf{return} $n_{t+1}$}
		\EndIf
		\State{$min\_energy \gets \texttt{UINT\_MAX}$}
        \State{$hist \gets \mathbb{H}[n_{t-1}]$} \Comment{Use the histogram at $n_{t-1}$}
		\ForAll{$\hat{n} \in hist.bins$} \Comment{Candidate values for worker counts}
		\State{$energy \gets 0$}
        \State{$n_{new} \gets 0$}
        \While{$\hat{n} > n_{curr} + n_{new} $} \Comment{Include amortized spin-up overhead}
        \State{$avg\_life \gets \mathbb{L}[n_{curr} + n_{new}] $}
        \State{$avg\_epochs \gets \lceil avg\_life / T_{s} \rceil $}
        \State{$energy \gets energy + B_{f} A_{f} / avg\_epochs $}
        \State{$n_{new} \gets n_{new} + 1$}
        \EndWhile
		\ForAll{$\{n, \rev{prob}\} \in hist$} \Comment{Iterate over the distribution}
		\If{$\hat{n} > n$} \Comment{Over-allocation}
		\State{$idle\_energy \gets (\hat{n} - n) I_{f}$}
		\State{$busy\_energy \gets n B_{f}$}
		\ElsIf{$\hat{n} < n$} \Comment{Under-allocation}
		\State{$idle\_energy \gets 0$}
		\State{$busy\_energy \gets \hat{n}B_{f} + (n - \hat{n}) S B_{c}$}
		\EndIf
		\State{$energy \gets energy + \rev{prob} \cdot (idle\_energy + busy\_energy)$}
		\EndFor
		\If{$energy < min\_energy$} \Comment{Pick $\hat{n}$ that minimizes $energy$}
		\State{$min\_energy \gets energy $}
		\State{$n_{t+1} \gets \hat{n}$}
		\EndIf
		\EndFor
		\State{\textbf{return} $n_{t+1}$}
		\EndProcedure
	\end{algorithmic}
\end{algorithm}

\Cref{alg:predict_num_fpgas} uses the histogram at $\mathbb{H}[n_{t-1}]$ to predict $n_{t+1}$ such that the expected energy consumption in that interval is minimised. 
If no histogram exists at $\mathbb{H}[n_{t-1}]$ (i.e., we are seeing that worker count value for the first time), the predictor maintains the same allocation as in the \rev{previous} interval.
If a histogram exists, the algorithm iterates through candidate values for $n_{t+1}$ taken from the histogram bins (or the range of values in between). 
It finds the $n_{t+1}$ that minimizes the expected energy usage, defined as the weighted sum of busy and idle energy usage across the entire conditional distribution. 
\rev{If the candidate value is higher than the currently allocated number of FPGAs, $n_{curr}$, the algorithm also includes the amortized spin-up energy for one interval, given the expected worker lifetime from $\mathbb{L}$.}
For each histogram entry (i.e., an expected worker count and its \rev{occurrence probability}),
if the candidate count is an over-allocation, it incurs busy energy overhead from the FPGAs used and idle energy overhead from the extra FPGAs; if it is an under-allocation, it incurs only busy energy overhead from all FPGAs and additional CPUs allocated on the dispatch path.
The algorithm sums the expected energy overhead across the distribution by weighting the contribution from each expected worker count by its occurrence \rev{probability}.
Finally, it selects the candidate value that minimizes \rev{the overall expected energy usage}.
To \rev{reduce} overhead, we cache results and lazily recalculate only if the histogram is updated.


\subsection{Request Dispatch and Fast Allocation}
\label{sec:energy_efficient_dispatch}


\Cref{alg:dispatch_algorithm} describes \sys's request dispatch strategy. 
\sys processes requests in deadline order to help meet latency requirements.
It estimates request sizes and tracks worker completion times to ensure that requests are only assigned to workers if they can finish on time. 
If none of the available workers can meet the deadline, the dispatcher spins up a new CPU worker and assigns it the pending request~\autocite{shahrad2020_Serverless,openfaas_OpenFaaS,cloudflare_Eliminating}.

\begin{algorithm}
	\footnotesize
	\caption{Efficient-first request dispatch algorithm.}
	\label{alg:dispatch_algorithm}
	\begin{algorithmic}[1]
		\Procedure{Dispatcher}{}
		\State{$Q \gets \textrm{request queue, ordered by deadline}$}
		\ForAll{$r \in Q$}
		\State{$w \gets \Call{FindAvailableWorker}{r}$}
		\If{$w == null$}
		\State{Spin up new CPU worker $c$ and assign $r$ to it}
		\Else
		\State{Assign $r$ to $w$}
		\EndIf
		\EndFor
		\State{\Call{Wait}{} until new requests arrive and repeat}
		\EndProcedure
		\Function{FindAvailableWorker}{$r$}
		\State{$\beta_y \gets \textrm{busy workers list, ordered by decr. load}$}
		\State{$\iota_y \gets \textrm{idle workers list, ordered by incr. time spent idle}$}
		\State{$\alpha_y \gets \textrm{list of workers being allocated, ordered by decr. queued load}$}
		\ForAll{$y \in \{f, c\}$}  \Comment{Efficient worker order (FPGA, CPU)}
		\ForAll{$w \in \{\beta_y, \iota_y, \alpha_y\}$}
		\If{\Call{CanMeetDeadline}{$w, r$}} \label{alg:can_meet_deadline}
		\State{\textbf{return} $w$}
		\EndIf
		\EndFor
		\EndFor
		\State{\textbf{return} $null$}
		\EndFunction
	\end{algorithmic}
\end{algorithm}

Request dispatching also impacts energy usage in hybrid platforms because it controls: 
(1)~whether to run requests on FPGAs or CPUs, and
(2)~whether idle workers can be reclaimed quickly.
Our dispatcher extends AutoScale's index-based dispatching scheme to hybrid workers~\autocite{gandhi2012_AutoScale}. 
Specifically, it dispatches requests in worker efficiency order as long as the target workers are expected to meet the deadline.
Among a pool of workers of the same type, it prefers, in order, workers that 
(1)~are busiest (i.e., have the highest load),
(2)~have been idle the least amount of time, and
(3)~have the highest queued load if they are currently being spun up. 
This policy maximizes efficiency by coalescing work on to workers that will be deallocated furthest in the future, allowing other workers to be reclaimed as soon as possible. 




\subsection{Trading Off Energy and Cost}

\rev{
A cost-optimized variant of \sys differs in two key ways: 
(1)~it uses a cost-based breakeven threshold in~\Cref{alg:per_interval_alloc}, given by $T_{b} = \frac{T_{s}C_{f}}{S C_{c}}$, and
(2)~it minimizes the overall cost rather than the energy usage when considering worker allocations in~\Cref{alg:predict_num_fpgas}.

\sys trades off energy and cost by minimizing configurable weighted sums of expected energy and cost.
In general, over-allocating FPGAs greatly increases costs but only slightly increases energy usage since FPGAs are expensive but consume low power; under-allocating FPGAs greatly increases energy usage but only slightly increases costs since the gap must be serviced by inefficient but cheap CPUs.
Operators can thus further customize the predictor based on their desired objectives.
}

\subsection{Limitations and Discussion}

\sys assumes that adequate workers are available to scale up and meet incoming load; it does not optimize for scarce resources.
\rev{Further, the calculation of the energy-optimal number of FPGAs in~\Cref{alg:predict_num_fpgas} ignores request deadlines.}
Deadline-aware FPGA allocations could make \sys more energy efficient (e.g., fewer FPGAs can handle the same load if the deadline is sufficiently long); however, for our target latency-sensitive applications, deadlines are typically very tight. 
We leave these extensions for future work.

Our description of \sys's dispatcher also assumes known request execution durations. 
This is a simplification.
Some domains like machine learning inference have very stable request sizes and these could be profiled in advance~\autocite{gujarati2020_Serving}.
If request sizes are not known, we can rely on online request size estimators from prior work, since they are shown to be effective for latency-sensitive applications~\autocite{chou2019mudpm,kasture2015rubik,iverson1999statistical}. 
Note that if request sizes come from a probability distribution, \sys can only meet probabilistic service-level objectives (SLOs) rather than hard deadlines~\autocite{mogul2019_Nines}. 

For simplicity of exposition, we have described \sys given existing differences between FPGAs and CPUs (e.g., vast gap in spin-up times). However, it can be generalized to arbitrary workers such as GPUs and other accelerators based on their characteristics.

\section{Evaluation}
\label{sec:evaluation}

Using \texttt{\sys}, we compare the trade-offs between hybrid and homogeneous platforms across a large set of workload and worker configurations. We show that:



\begin{itemize}[nosep]
	\item \sys uses up to 1.53$\times$ lower energy and costs up to 2.6$\times$ less than FPGA-only platforms on production cloud workloads under plausible application and worker configurations.
    \item \sys variants can effectively trade off energy efficiency for cost; they outperform existing cost-optimized hybrid schedulers since they prioritize FPGA allocations and provide a request dispatch policy that maximizes FPGA utilization.
	\item Hybrid platforms can harness the efficiency benefits of FPGAs at a reasonable cost across a wide spectrum of configurations. They are especially effective for applications with short requests and moderate-to-high workload burstiness.
\end{itemize}


\subsection{Methodology}
\label{sec:methodology}

We implement a discrete-event simulator in Cython and C++ to evaluate hybrid and homogeneous platforms. This section describes how we set up experiments and configure our simulator.

\begin{table}
	\footnotesize
	\centering
	\begin{tabular}{@{}ccc@{}}
		\toprule
		\textbf{}         & \textbf{CPU Worker}         & \textbf{FPGA Worker}                               \\ \midrule
		Spin-up latency   & 5ms                         & \emph{1s}, 10s, \emph{60s, 100s}                 \\
		Spin-down latency & 5ms                         & 100ms                                              \\
		Relative speedup  & 1$\times$                   & \emph{1$\times$}, \rev{2$\times$, \emph{4$\times$}} \\
		Busy power        & \rev{150W}                  & \emph{25W}, 50W, \emph{100W}                       \\
		Idle power        & \emph{10W}, 30W, \emph{50W} & \emph{10W}, 20W, \emph{30W}                        \\
		Prorated cost & \$0.668/hr                  & \$0.982/hr                                         \\
		\bottomrule
	\end{tabular}
	\caption{Worker configurations. Non-italicized values are defaults.}
	\label{tab:eval_defaults}
\end{table}

\medskip
\noindent\textbf{Workers.} CPU and FPGA workers are parameterized by~\Cref{tab:eval_defaults}, based on our survey in~\Cref{sec:cpus_vs_fpgas}. 
We assume that workers draw busy power during spin up and spin down.
Our defaults represent current trends: FPGA workers spin-up slowly, use less energy, and cost more per instance than CPUs.
We note that although these are plausible parameters, for any specific application, the cost and efficiency trade-offs between workers will vary.

\begin{table}
	\footnotesize
	\centering
	\begin{tabular}{@{}ccc@{}}
		\toprule
		\textbf{Request Size}         & \textbf{Azure Functions} & \textbf{Alibaba $\mu$services} \\ \midrule
		Short (10\ms--100\ms)     &  13                             & 99                       \\
		Medium (100\ms--1\s)      & 101                             & 31                      \\
		Long (1\s--10\s)          & 241                             & N/A                      \\
		\bottomrule
	\end{tabular}
	\caption{Number of heavy-demand applications evaluated from the production traces, according to our request size buckets.}
	\label{tab:traces}
    \vspace{-\baselineskip}
\end{table}

\medskip
\noindent\textbf{Workloads.} We evaluate \texttt{\sys} on both production and synthetic traces.
While no public data sets capture FPGA usage, we repurpose two CPU-based production traces shown in~\Cref{tab:traces}.
The Azure trace consists of serverless function invocations on their public cloud platform~\autocite{shahrad2020_Serverless}.
The Alibaba trace consists of remote procedure call invocations of production microservices~\autocite{luo2021_Characterizing}. 
Both data sets show very skewed compute demand: fewer than 25\% of the applications require more than one worker at any point, but they constitute over 94\% of the compute demand.
We focus on this subset since the other applications have low load and are best run on CPUs. 
Both datasets provide request sizes and per-minute request arrival rates.
We use the request rates to generate two-hour traces with time-varying Poisson interarrivals, assuming that the rates change linearly within each minute.
We evaluate applications that fit into our request size buckets spanning short (10\ms--100\ms), medium (100\ms--1\s), and long (1\s--10\s) requests.

In addition, we generate synthetic traces for sensitivity analysis using a three step process. 
First, we randomly sample the request size (for a CPU worker) from one of the three aforementioned buckets.
Based on the selected request size, we next generate per-minute request arrival rates based on a self-similar distribution such that, on average, 100 CPU workers are needed to serve the load. 
Self-similar rate traces are generated using the b-model~\autocite{wang2002_Data}, where the bias parameter captures burstiness ranging from 0.5 (uniform load) to 0.75 (highly variable).
Finally, we use these rate-based traces to generate time-varying Poisson interarrivals like above. 
Unless specified, we use two hour, short request size traces with a burstiness of 0.6.

\medskip
\noindent\textbf{Baselines.} We compare energy-optimized (\texttt{\sysE}), cost-optimized (\texttt{\sysC}), and balanced (\texttt{\sysB}) variants of \sys with the following homogeneous baselines:
(1)~\texttt{CPU-dynamic} is a CPU-only reactive scheduler based on serverless frameworks and AutoScale~\autocite{gandhi2012_AutoScale} that uses fast spin ups to handle bursty load (equivalent to \texttt{\sys} with only CPU workers). 
(2)~\texttt{FPGA-static} has perfect workload information and pre-allocates exactly enough FPGAs to handle the peak load; it represents the best-case statically provisioned FPGA-only platform~\autocite{putnam2014_Reconfigurable, alibaba_Deep, amazonwebservices_EC2}.
(3)~\texttt{FPGA-dynamic} is a FPGA-only reactive scheduler that maintains a fixed excess headroom above the minimum FPGAs needed to meet current load~\autocite{gandhi2012_AutoScale,microsoftazure_Azure,amazon_ecs_capacityscaling}. 
For each trace, \texttt{FPGA-dynamic} allocates the least headroom that meets request deadlines based on an integer multiple of the maximum difference in known request rates between consecutive intervals.

We also compare \sys to MArk~\autocite{zhang2019_MArk}, a state-of-the-art, cost-optimized scheduler for CPU-GPU hybrids.
Like \sys, MArk combines predictive and reactive worker management and it relies on request size estimations to meet deadlines (otherwise, it provides probabilistic SLOs~\autocite{mogul2019_Nines}).
MArk differs from \sys in the following key ways. MArk:
(1)~is only cost optimized,
(2)~uses a slow, data-intensive LSTM model for predictions~\autocite{hochreiter1997long},
(3)~predicts fine-grained request rates over multiple intervals (a challenging prediction problem) rather than coarse-grained worker counts for one interval, 
and
(4)~uses round-robin request dispatch, which is inefficient and makes it harder to reclaim workers when workload intensity drops.
MArk's performance crucially depends on the accuracy of its predictor, which was not the target of their work~\autocite{zhang2019_MArk}.
To give it the benefit of doubt and to focus on the key differences between the two scheduling approaches, we evaluate an idealized version of MArk, \texttt{MArk-ideal}, which has perfect workload predictions up to two intervals into the future. 
We similarly also evaluate \texttt{\sysE-ideal} and \texttt{\sysC-ideal}, which both have perfect workload predictions for the subsequent interval, ignoring spin-up overhead accounting.

All schedulers keep workers idle for as long as the allocation duration before spinning them down in anticipation of future requests~\autocite{gandhi2012_AutoScale,shahrad2020_Serverless,karlin1994competitive}.
Periodic schedulers use scheduling intervals of length equal to the FPGA spin-up time. 
We run non-ideal \sys variants without warming up their predictors; in production deployments, the predictors can learn the workload distribution over time, yielding better results. 

\medskip
\noindent\textbf{Metrics.} 
We report the energy efficiency and cost of each scheduler relative to an idealized, best-case FPGA-only platform with default parameters that incurs only compute costs (i.e., zero idling and spin-up overheads). 
We also ensure that requests meet deadlines, defined to be 10$\times$ the request size.
Synthetic workload experiment results are averaged across ten randomly generated traces, whereas production workloads are run once. 
We present a subset of results.


\begin{table}
\begin{subtable}{\columnwidth}
\footnotesize
\centering
\begin{tabular}{@{}c|cc|cc@{}}
\toprule
\textbf{Short Reqs.} & \multicolumn{2}{c|}{\textbf{Azure Functions (13 apps)}}                                                                                                                                                & \multicolumn{2}{c}{\textbf{Alibaba $\mu$services (99 apps)}}                                                                                                                                               \\ \midrule
\textbf{Scheduler}    & \textbf{\begin{tabular}[c]{@{}c@{}}Energy\\ Efficiency\end{tabular}} & \textbf{\begin{tabular}[c]{@{}c@{}}Relative\\ Cost\end{tabular}} & \textbf{\begin{tabular}[c]{@{}c@{}}Energy\\ Efficiency\end{tabular}} & \textbf{\begin{tabular}[c]{@{}c@{}}Relative\\ Cost\end{tabular}} \\ \midrule
CPU-dynamic                            & 16.5\%                                                          & 1.35$\times$                                                                                                                    & 16.6\%                                                          & 1.34$\times$                                                                                                                    \\
FPGA-static                         & 54.4\%                                                         & 3.08$\times$                                                                                                                    & 79.4\%                                                         & 1.64$\times$                                                                                                                    \\
FPGA-dynamic                           & 56.2\%                                                         & 2.87$\times$                                                                                                                    & 77.7\%                                                         & 1.68$\times$                                                                                                                    \\
MArk-ideal                              & 36.2\%                                                         & 1.41$\times$                                                                                                                   & 62.2\%                                                         & 1.14$\times$                                                          \\
\sysC                              & 73.5\%                                                         & 1.17$\times$                                                                                                               & 84.8\%                                                         & 1.08$\times$                                                          \\
\sysB                              & 82.4\%                                                         & 1.23$\times$                                                                                                               & 91.9\%                                                         & 1.12$\times$                                                          \\
\sysE                              & 86.2\%                                                         & 1.34$\times$                                                                                                & 92.8\%                                                         & 1.15$\times$                                                                                   \\
\sysC-ideal                        & 76.6\%                                                         & 1.19$\times$                                                                                                           & 90\%                                                         & 1.1$\times$                                                                                          \\
\sysE-ideal                        & 87.2\%                                                         & 1.31$\times$                                                                                   & 93.1\%                                                         & 1.14$\times$                            \\
\bottomrule
\end{tabular}
\caption{Short request sizes.}
\end{subtable}\\
\begin{subtable}{\columnwidth}
\footnotesize
\centering
\begin{tabular}{@{}c|cc|cc@{}}
\toprule
\textbf{Medium Reqs.} & \multicolumn{2}{c|}{\textbf{Azure Functions (101 apps)}}                                                                                                                                                & \multicolumn{2}{c}{\textbf{Alibaba $\mu$services (31 apps)}}                                                                                                                                               \\ \midrule
\textbf{Scheduler}    & \textbf{\begin{tabular}[c]{@{}c@{}}Energy\\ Efficiency\end{tabular}} & \textbf{\begin{tabular}[c]{@{}c@{}}Relative\\ Cost\end{tabular}} & \textbf{\begin{tabular}[c]{@{}c@{}}Energy\\ Efficiency\end{tabular}} & \textbf{\begin{tabular}[c]{@{}c@{}}Relative\\ Cost\end{tabular}} \\ \midrule
CPU-dynamic           & 16.6\%                                                           & 1.33$\times$                                                                                                                 & 16.6\%                                                           & 1.34$\times$                                                                                                           \\
FPGA-static           & 72.2\%                                                         & 1.96$\times$                                                                                                            & 79.7\%                                                           & 1.63$\times$                                                                                                      \\
FPGA-dynamic          & 78.8\%                                                         & 1.63$\times$                                                                                                       & 77.6\%                                                         & 1.67$\times$                                                                                              \\
MArk-ideal            & 74.8\%                                                         & 1.11$\times$                                                                                                        & 62.2\%                                                         & 1.16$\times$                                                                                                        \\
\sysC                & 81.2\%                                                           & 1.05$\times$                                                                                                         & 79.5\%                                                         & 1.05$\times$                                                        \\
\sysB                & 87.6\%                                                           & 1.09$\times$                                                                                                         & 86.4\%                                                         & 1.07$\times$                                                        \\
\sysE                & 91\%                                                           & 1.17$\times$                                                                                                   & 92.1\%                                                         & 1.16$\times$                            \\
\sysC-ideal        & 92.4\%                                                         & 1.06$\times$                                                                                                             & 88.5\%                                                         & 1.06$\times$                                                                                                        \\ 
\sysE-ideal        & 94.4\%                                                         & 1.09$\times$                                                                                                      & 93\%                                                         & 1.13$\times$                                                                                                    \\ 
\bottomrule
\end{tabular}
\caption{Medium request sizes.}
\end{subtable}%
\caption{Energy efficiency, cost, and deadline misses for production traces with short and medium request sizes. Energy consumption and costs are aggregated across all applications and reported relative to an idealized FPGA-only platform.}
\label{tab:production_efficiency_cost}
\vspace{-\baselineskip}
\end{table}

\subsection{Production Workloads}

\Cref{tab:production_efficiency_cost} shows the overall energy efficiency and cost of various schedulers running the Azure and Alibaba traces under default configurations. 
We find that all \texttt{\sys} variants outperform homogeneous platforms across all metrics and traces. 
Specifically, for the Azure traces with short requests, \texttt{\sysE} costs about the same as \texttt{CPU-dynamic} but is 5.2$\times$ more energy efficient; further, it is is up to 1.53$\times$ more energy efficient and costs up to 2.14$\times$ lower than \texttt{FPGA-static} and \texttt{FPGA-dynamic}, thereby achieving the best of both worlds.
\texttt{\sysE}'s efficiency and cost improvements also apply to the Alibaba traces, but its relative benefit over FPGAs reduces due to a less bursty workload. Specifically, \texttt{\sysE} is 1.16$\times$ cheaper than \texttt{CPU-dynamic} and it is 1.16$\times$ more efficient and 1.43$\times$ cheaper than \texttt{FPGA-static}.
Benefits are also lower for medium request sizes. 
For example, on the Azure traces, \texttt{\sysE} is 1.15$\times$ more energy efficient and 1.4$\times$ cheaper than \texttt{FPGA-dynamic}.

Among hybrid schedulers, all \sys variants are strictly better than \texttt{MArk-ideal} despite the fact that they start with no prior workload information. 
Notably, although \texttt{MArk-ideal} and \texttt{\sysC} are both cost optimized, the latter provides 1.1--2$\times$ higher energy efficiency while costing 1.06--1.2$\times$ lower.
\texttt{\sysC-ideal} further improves upon \texttt{\sysC}'s benefits due to perfect predictions.
We find that \sys variants can appropriately trade-off energy efficiency for cost on all traces.
\texttt{\sysE} is up to 1.17$\times$ more energy efficient than \texttt{\sysC}, but costs up to 1.14$\times$ higher. \texttt{\sysB} achieves a middle ground on both metrics. 


We further note that \sys's simple prediction mechanism works reasonably well in practice.
\texttt{\sysE} and \texttt{\sysC} are consistently close to their ideal variants both in terms of energy efficiency and cost.
Intuitively, when FPGA workers are 6$\times$ more energy efficient than CPUs (our default configuration), executing only 1\% of the requests on CPUs would result in $\sim$5\% energy efficiency loss relative to an idealized FPGA-only platform, ignoring spin up and idle overheads.
Our result shows that for real workloads, \sys runs most of the requests on FPGAs and maximizes their utilization.

\begin{table}
\footnotesize
\centering
\begin{tabular}{@{}cccc@{}}
\toprule
\textbf{Trace (request size)}  & \textbf{Round Robin~\autocite{zhang2019_MArk}} & \textbf{Index Packing~\autocite{gandhi2012_AutoScale}} & \textbf{Spork} \\ \midrule
Azure (short)    & 60.6\%                 & 85.5\%             & 86.2\%                   \\
Azure (medium)   & 83.6\%               & 81.9\%             & 91\%                     \\
Azure (long)     & 85.6\%               & 84.2\%             & 89.9\%                   \\
Alibaba (short)  & 76.9\%               & 88.8\%               & 92.8\%                   \\
Alibaba (medium) & 85.7\%                 & 91.3\%             & 92.1\%                   \\ \bottomrule
\end{tabular}
\caption{Energy efficiency impact of different dispatch policies on production workloads under \texttt{\sysE}'s worker allocation logic.}
\label{tab:balancers}
\vspace{-\baselineskip}
\end{table}

\subsection{Understanding \sys's Efficacy}
\label{sec:scheduling_efficacy}

A key reason for \sys's advantage over MArk is due to its request dispatch policy, designed specifically for hybrid platforms.
\Cref{tab:balancers} compares the energy efficiency of \sys's dispatcher with MArk's round robin~\autocite{zhang2019_MArk} and AutoScale's index packing dispatchers~\autocite{gandhi2012_AutoScale} on production workloads. 
For consistency, all configurations are run with \texttt{\sysE}'s worker allocation logic.
We find that \sys's dispatcher is up to 42\% more energy efficient than round robin and up to 11\% more than index-based dispatching.
Both prior policies suffer from drawbacks:
round robin evenly distributes requests to both CPUs and FPGAs and it rarely lets workers idle;
index packing often dispatches to busy but inefficient CPU workers over idle FPGAs.
In contrast, \sys's dispatcher maximizes FPGA utilization and helps reclaim workers quickly, yielding high efficiency.

\begin{figure}
	\centering
	\includegraphics[width=3.375in]{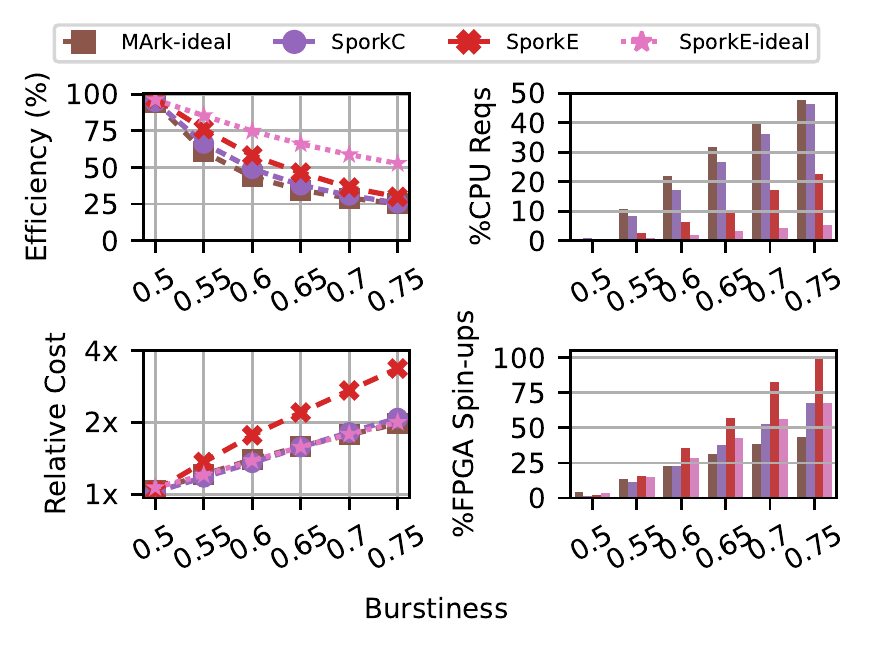}
	\caption{Energy efficiency and cost trade-offs between \sys and MArk~\autocite{zhang2019_MArk} under varying burstiness and a 60\s FPGA spin up (left). Percentage of requests executed on CPU workers and the total number of FPGA spin ups normalized to tha maximum under each scheduler (right). \texttt{\sys-ideal} and \texttt{MArk-ideal} have perfect predictions.}
	\label{fig:ideal_vs_real}
    \vspace{-\baselineskip}
\end{figure}



We next use synthetic traces to understand the impact of prediction error.
\Cref{fig:ideal_vs_real} (left) evaluates the energy efficiency and cost trade-offs between \texttt{MArk-ideal} and \sys variants with increasing burstiness.  
Higher burstiness renders \sys's predictions less effective.
Longer intervals also make it harder for \sys to accurately predict, so we set the scheduling interval length (and FPGA spin-up) to 60s.
We find that:
(1)~\texttt{\sysC} and \texttt{\sysE-ideal} provide up to 1.13$\times$ and 2.1$\times$ higher energy efficiency than \texttt{MArk-ideal} while costing about the same,
(2)~\texttt{\sysE} outperforms \texttt{MArk-ideal} by up to 1.34$\times$ in terms of energy efficiency but incurs a 1.38$\times$ higher cost at moderate burstiness, and
(3)~\texttt{\sysE} differs from \texttt{\sysE-ideal} by up to 1.75$\times$ in energy efficiency and cost at high burstiness.

To investigate why, we plot for each scheduler the percentage of total requests executed on CPUs and the number of FPGAs allocated (normalized to the maximum) over the entire workload in~\Cref{fig:ideal_vs_real} (right).
We find that \texttt{MArk-ideal} and \texttt{\sysC} have lower costs since they allocate fewer FPGAs and run up to half of the requests on CPUs; however, heavily using CPUs reduces their energy efficiency.
\texttt{\sysE} runs only up to 22\% of the requests on CPUs providing higher energy efficiency; however, mispredictions cause it to allocate almost 2.3$\times$ more FPGAs than \texttt{MArk-ideal} leading to higher cost.
\texttt{\sysE-ideal}'s perfect predictions further enhance the efficiency advantage; it executes less than $\sim$5\% of the requests on CPUs to handle transient bursts. 
Although \texttt{\sysE-ideal} allocates up to 1.5$\times$ more FPGAs than \texttt{MArk-ideal}, it has similar costs overall since these allocations are right-sized and coupled with a dispatcher that maximizes their utilization.

\begin{figure}
	\centering
	\includegraphics[width=3.375in]{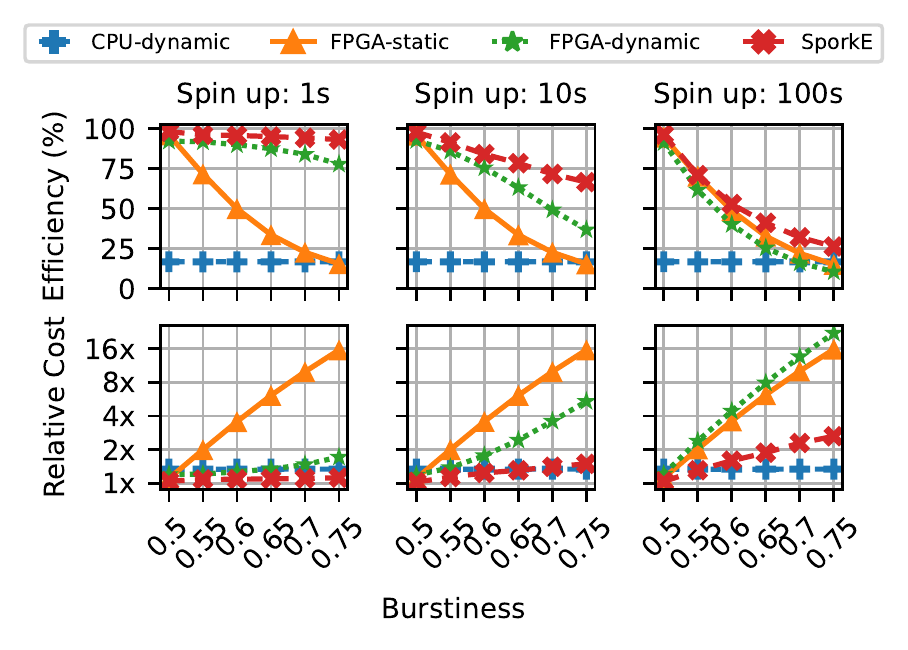}
	\caption{Sensitivity to workload burstiness and FPGA spin up costs. Energy efficiency and cost are reported relative to an idealized FPGA-only baseline with default parameters from~\Cref{tab:eval_defaults}. Each data point is averaged across 10 trace runs.}
	\label{fig:burstiness_fpgaalloc}
    \vspace{-\baselineskip}
\end{figure}

\subsection{Sensitivity Analysis}
\label{sec:hybrid_vs_homogeneous}

We quantify the sensitivity to our assumptions by comparing \sysE with homogeneous platforms across several configurations.

\medskip
\noindent\textbf{Workload Burstiness and FPGA Spin-up Costs.}
\Cref{fig:burstiness_fpgaalloc} shows the energy efficiency and cost impact of varying workload burstiness and FPGA spin-up times.
We observe that:
(1)~\texttt{CPU-reactive} is cost effective at higher burstiness, while FPGA-only platforms are generally energy efficient despite overprovisioning,
(2)~\texttt{FPGA-dynamic} loses its advantage over \texttt{FPGA-static} with increasing FPGA spin-up times,
(3)~\texttt{\sysE} is better than or equivalent to others in terms of energy efficiency while being much cheaper than FPGA-only,
(4)~Higher burstiness improves \texttt{\sysE}'s efficiency and cost benefits over FPGA-only platforms, and
(5)~Longer FPGA spin ups reduce \texttt{\sysE}'s energy efficiency benefits, but they increase its cost advantage over \texttt{FPGA-dynamic}.
We elaborate on these below.

Among homogeneous platforms, \texttt{CPU-dynamic} is cost effective compared to FPGA-only platforms at moderate-to-high burstiness---up to 11.8$\times$ cheaper than \texttt{FPGA-static}.
In contrast, \texttt{FPGA-static} and \texttt{FPGA-dynamic} are more energy efficient despite overprovisioning.
Their advantage vanishes with higher burstiness and spin-up costs, where using only CPUs is preferable.
These results match the trends we observe in~\Cref{sec:motivation}.
Among FPGA-only, \texttt{FPGA-dynamic} significantly outperforms \texttt{FPGA-static} when spin up times are small.
With long spin ups, \texttt{FPGA-dynamic} performs worse since it needs a larger headroom to handle bursts. 
Like in traditional autoscaling systems~\autocite{nguyen2013agile,microsoftazure_Azure,amazon_ecs_capacityscaling}, \texttt{FPGA-dynamic} maintains this headroom as load fluctuates, which causes it to sometimes allocate more workers than \texttt{FPGA-static}.
In contrast, \texttt{FPGA-static} only incurs a minor one-time spin-up cost; it also benefits from precise provisioning due to perfect workload knowledge.

\texttt{\sysE} provides significant energy efficiency and cost benefits over homogeneous platforms without prior workload knowledge (barring request sizes). 
Even at high burstiness, \texttt{\sysE} is up to 1.2$\times$ cheaper and 5.6$\times$ more energy efficient than \texttt{CPU-reactive}, but the cost advantage is lost under long spin ups.
\texttt{\sysE} is also up to 2.5$\times$ more efficient (8.4$\times$ cheaper) than \texttt{FPGA-dynamic} and up 6$\times$ more efficient (13.9$\times$ cheaper) than \texttt{FPGA-static}.

\texttt{\sysE}'s improvement over FPGA-only increases with burstiness.
For example, with 10\s spin ups at moderate burstiness, \texttt{\sysE} is 1.1$\times$ more efficient and 1.4$\times$ cheaper than \texttt{FPGA-dynamic}; at high burstiness, this increases to 1.8$\times$ higher efficiency and 3.6$\times$ lower cost. 
FPGA-only platforms suffer since they allocate expensive FPGAs that are only briefly utilized.

Like \texttt{FPGA-dynamic}, \texttt{\sysE}'s efficiency benefits reduce as the FPGA spin-up times increase because:
(1)~spin-up costs add up,
(2)~there is a higher probability of incurring request bursts over a long interval, and
(3)~predictions become harder, especially since there are fewer samples for the predictor to learn from.
Yet, \texttt{\sysE} becomes much cheaper (up to $8.4\times$) than \texttt{FPGA-dynamic} with longer spin ups, since the latter is worse off due to provisioning a large FPGA headroom.
Compared to \texttt{FPGA-static} with a 100\s spin up at moderate burstiness, \texttt{\sysE} is only 1.08$\times$ more energy efficient but it still remains 2.3$\times$ cheaper.

\begin{figure}
	\centering
	\includegraphics[width=3.375in]{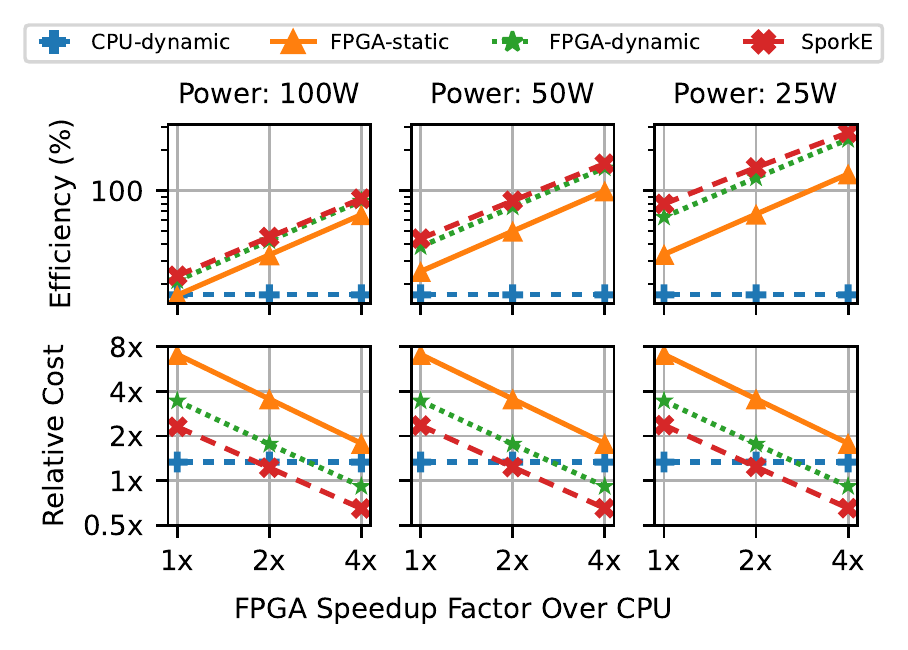}
	\caption{Sensitivity to FPGA performance and busy power draw. Note that both axes are in log scale.}
	\label{fig:platform_efficiency}
    \vspace{-\baselineskip}
\end{figure}

\medskip
\noindent\textbf{Worker Efficiency.} \Cref{fig:platform_efficiency} shows the energy efficiency and cost impact as FPGAs become faster and more power efficient than CPUs at serving requests.\footnote{In our model, power efficiency improvements do not affect cost since we use fixed cloud prices. Trends remain similar even after including operational costs, based on a total cost of ownership (TCO) analysis~\autocite{barroso2018_Datacenter}.}
In all cases, \texttt{\sysE} retains its cost advantage over both FPGA-only alternatives, while providing similar or better energy efficiency.
We make the following additional observations. 

First, FPGA processing speedups provide a greater advantage to FPGA-only platforms than to hybrid platforms. 
A 4$\times$ FPGA speedup offers a near-linear improvement in both energy efficiency and cost for FPGA-only platforms.
\texttt{\sysE} sees less improvement (3.4--3.7$\times$) since it runs some of the workload on CPU workers. 
Higher speedups make \texttt{\sysE}'s mispredictions relatively more expensive.
They can even make FPGA-only platforms strictly preferable over CPU-only due to the significant cost reduction---for example, at a 4$\times$ speedup, \texttt{FPGA-dynamic} is up to 1.47$\times$ cheaper than \texttt{CPU-dynamic}. 

Second, optimizing the busy power efficiency of FPGAs has diminishing returns for FPGA-only platforms since idle power starts to dominate.
For example, a 4$\times$ reduction in FPGA busy power yields only $\sim$2$\times$ lower energy usage for \texttt{FPGA-static}, whereas \texttt{\sysE}'s energy usage reduces by over 3$\times$ since it better utilizes FPGAs.
Idling accounts for 33\% of \texttt{FPGA-static}'s overall energy consumption when the FPGA busy power is 100W; this ratio increases to over 66\% at 25W.
For \texttt{\sysE}, the idling energy contribution goes from only 4\% to 13\%.
As busy power draw is optimized, the efficiency gap between \texttt{\sysE} and \texttt{FPGA-dynamic} also widens (from 1.04$\times$ to 1.12$\times$) since the latter increasingly incurs idling overheads.
In general, FPGA speed ups help more than power draw optimizations since fewer workers can be used to serve the same workload.


\begin{figure}
	\centering
	\includegraphics[width=3.375in]{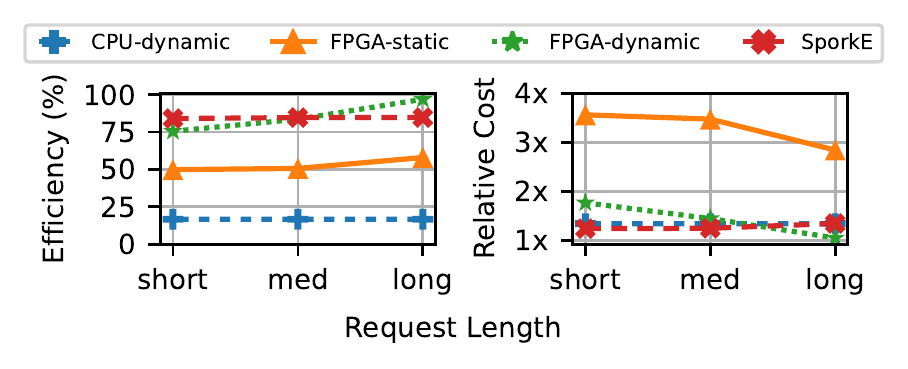}
	\caption{Sensitivity to request sizes. We consider
		short (10\ms--100\ms), medium (100\ms--1\s), and long (1\s--10\s) requests. Deadlines are 10$\times$ the request size.}
	\label{fig:request_length}
    \vspace{-\baselineskip}
\end{figure}

\medskip
\noindent\textbf{Request Sizes.}
~\Cref{fig:request_length} examines the impact of varying application request sizes (and correspondingly, deadlines). 
Longer requests and deadlines substantially benefit FPGA-only platforms since they improve worker utilization and diminish the impact of slow spin-up times.
\texttt{FPGA-static}'s energy efficiency increases by up to 16\% with long requests since it needs fewer workers to service the same load.
\texttt{FPGA-dynamic}'s energy efficiency goes up by 1.3$\times$ and cost down by 1.7$\times$ since it needs to maintain a smaller headroom. 
In fact, for requests that are over 1\s long, no headroom is necessary since the deadlines exceed the FPGA spin-up time. 
In such cases, a purely reactive scheduler works well since it can automatically adapt to load while having sufficient leeway to meet deadlines.
\texttt{\sys}'s benefits decline with longer requests because it does not consider deadlines for FPGA allocation (\Cref{sec:fpga_allocation}).

\section{Related Work}
\label{sec:related}

\rev{To our knowledge, we are the first to
show that hybrid FPGA-CPU platforms can harness the energy efficiency benefits of FPGAs for latency-sensitive and bursty applications at a low cost and build a hybrid scheduler that can trade off the two metrics.}

\smallskip
\noindent\textbf{Datacenter FPGAs.}
FPGAs are traditionally bundled under the virtual machine abstraction and managed as peripherals~\autocite{amazonwebservices_EC2,alibaba_Deep,khawaja2018_Sharing,zha2020_Virtualizing,ma2020_Hypervisor}.
Several efforts elevate them to first-class compute and manage them directly with datacenter orchestrators like OpenStack~\autocite{byma2014_FPGAs,tarafdar2019_Building,eskandari2019_Modular,tarafdar2017_Enabling,ringlein2019_System,weerasinghe2016_Disaggregated,weerasinghe2016_Networkattached,abel2017_FPGA},
Others customize FPGA use for specific applications such as inference serving~\autocite{azuremachinelearning_Deploy,hoover2021_1stCLaaS,chung2018_Serving}, key-value stores~\autocite{istvan2017_Caribou,lavasani2014_FPGAbased}, or big data processing~\autocite{huang2016_Programming}.
While these works enable diverse FPGA use cases in datacenters, they do not support efficiently sharing them among latency-sensitive and bursty applications, necessitating overprovisioning.
More recently, FPGAs have been incorporated into serverless frameworks, enabling dynamic management on shared worker pools~\autocite{bacis2020_BlastFunction,ojika2019_FaaM,hoover2021_1stCLaaS,du2022_Serverless,inaccel_FPGA}. 
Perhaps the closest to our design is Molecule~\autocite{du2022_Serverless}, which provides a serverless abstraction on CPUs and FPGAs. 
Molecule optimizes FPGA spin-ups down to a few seconds for small applications and it helps share the FPGA across multiple instances to improve efficiency. 
However, it does not provide a scheduler to interchangeably run requests on different backends that optimizes the overall efficiency and cost. 

\smallskip
\noindent\textbf{Hybrid Computing.}
Prior work has primarily focused on hybrid GPU-CPU platforms, which have different trade-offs than FPGA-CPU platforms.
Most efforts target machine learning inference serving and attempt to optimize cost and throughput while meeting latency deadlines~\autocite{zhang2019_MArk, romero2019_INFaaS,yang2022_INFless,nvidia_Triton}. 
They often use domain-specific techniques like request batching and model variants to improve performance.
AlloX targets batch jobs like deep learning training and it provides a scheduling algorithm to optimize performance while providing fairness guarantees~\autocite{le2020_AlloX}.
However, none of these works consider energy efficiency; they also do not explore trade-offs of using hybrid platforms for latency-sensitive and bursty applications under diverse workload configurations.


\smallskip
\noindent\textbf{Interactive Scheduling.}
Schedulers can be broadly classified as reactive or predictive.
Reactive schedulers dynamically respond to demand and perform well when request deadlines are longer than worker spin-up times~\autocite{gandhi2012_AutoScale,suresh2019_FnSched,tariq2020_Sequoia,openfaas_OpenFaaS}; most serverless schedulers fall into this category.
Predictive schedulers attempt to proactively right-size worker allocations to meet application SLOs~\autocite{gujarati2017_Swayam,kannan2019_GrandSLAm,zhang2019_MArk,singhvi2019_Archipelago}. 
\sys incorporates both aspects and proposes a new prediction mechanism to improve efficiency~\autocite{gandhi2011minimizing,zhang2019_MArk}.
Depending on the optimization metric, request dispatch mechanisms in existing schedulers either evenly distribute load to improve utilization and tail latency~\autocite{zhang2019_MArk,ousterhout2019_Shenango} or consolidate load for higher efficiency~\autocite{suresh2019_FnSched,gandhi2012_AutoScale}. 
We extend the latter to hybrid platforms.
Some serverless schedulers attempt to minimize the impact of cold starts (i.e., spin ups) through prediction~\autocite{shahrad2020_Serverless,roy2022_IceBreaker,yang2022_INFless} and caching~\autocite{fuerst2021_FaasCache}.
However, cold start reduction targets infrequently used applications---these would be inefficient to run on FPGAs.

\section{Conclusion}
\label{sec:conclusion}

We propose a hybrid FPGA-CPU computing framework to take advantage of FPGA energy efficiency while reducing costs for latency-sensitive and bursty applications.
To achieve these benefits, \rev{we devise \sys, an efficient hybrid scheduler that optimizes for and can trade off energy efficiency and cost.
\sys includes}:
(1)~a lightweight predictor that periodically allocates FPGAs, and 
(2)~a heterogeneity-aware request dispatcher that can exploit fast CPU spin ups.
Our scheduler captures key differences between FPGA and CPU workers in terms of their efficiency, costs, and spin-up times. 
We evaluate \sys using production and synthetic traces under a large set of workload and worker configurations. 
Our results on cloud workloads with \rev{tight deadlines} show that \rev{energy-optimized \sys uses up to 1.53$\times$ lower energy and costs 2.14$\times$ less than FPGA-only platforms.
Compared to an idealized version of MArk, a cost-optimized hybrid scheduler, energy-optimized \sys provides up to 1.2--2.4$\times$ higher energy efficiency at similar cost.}

\begin{acks}
This work is supported by NSF grant CNS-2104548 and research grants by VMware and Cisco.
\end{acks}

\printbibliography

@inproceedings{abel2017_FPGA,
  title = {An {{FPGA Platform}} for {{Hyperscalers}}},
  author = {Abel, Francois and Weerasinghe, Jagath and Hagleitner, Christoph and Weiss, Beat and Paredes, Stephan},
  date = {2017},
  url = {http://ieeexplore.ieee.org/document/8071053/},
  eventtitle = {Annual Symposium on High-{{Performance Interconnects}} ({{HOTI}})}
}

@inproceedings{agache2020_Firecracker,
  title = {Firecracker: {{Lightweight Virtualization}} for {{Serverless Applications}}},
  author = {Agache, Alexandru and Brooker, Marc and Iordache, Alexandra and Liguori, Anthony and Neugebauer, Rolf and Piwonka, Phil and Popa, Diana-Maria},
  date = {2020},
  url = {https://www.usenix.org/conference/nsdi20/presentation/agache},
  urldate = {2020-11-28},
  eventtitle = {{{USENIX Symposium on Networked Systems Design}} and {{Implementation}} ({{NSDI}})}
}

@online{alibaba_Deep,
  title = {Deep {{Dive}} into {{Alibaba Cloud F3 FPGA}} as a {{Service Instances}}},
  author = {Alibaba Cloud},
  url = {https://www.alibabacloud.com/blog/deep-dive-into-alibaba-cloud-f3-fpga-as-a-service-instances_594057},
  urldate = {2021-05-13},
}

@online{amazonwebservices_EC2,
  title = {{{EC2 F1 Instances}}},
  author = {Amazon Web Services},
  url = {https://aws.amazon.com/ec2/instance-types/f1/},
  urldate = {2021-05-13}
}

@online{alibaba_pricing,
  title = {{{Pricing Calculator}}},
  author = {Alibaba Cloud},
  url = {https://www.alibabacloud.com/pricing-calculator},
  urldate = {2023-01-13}
}

@online{apache_OpenWhisk,
  title = {{{OpenWhisk}}},
  author = {Apache},
  url = {https://openwhisk.apache.org/},
  urldate = {2021-05-14}
}

@online{AWS_EC2_pricing,
  title = {{{EC2 On-Demand Instance Pricing}}},
  author = {Amazon Web Services},
  url = {https://aws.amazon.com/ec2/pricing/on-demand/},
  urldate = {2022-04-30},
}

@online{azuremachinelearning_Deploy,
  title = {Deploy {{ML}} Models to {{FPGAs}}},
  author = {Microsoft Azure},
  url = {https://docs.microsoft.com/en-us/azure/machine-learning/how-to-deploy-fpga-web-service},
  urldate = {2021-05-13}
}

@inproceedings{bacis2020_BlastFunction,
  title = {{{BlastFunction}}: {{An FPGA-as-a-Service System}} for {{Accelerated Serverless Computing}}},
  author = {Bacis, Marco and Brondolin, Rolando and Santambrogio, Marco},
  date = {2020},
  url = {https://ieeexplore.ieee.org/document/9116333/},
  eventtitle = {Design, {{Automation}}, and {{Test}} in {{Europe}} ({{DATE}})}
}

@article{barroso2018_Datacenter,
  title = {The {{Datacenter}} as a {{Computer}}: {{Designing Warehouse-Scale Machines}}},
  author = {Barroso, Luiz André and Hölzle, Urs and Ranganathan, Parthasarathy},
  date = {2018},
  journaltitle = {Synthesis Lectures on Computer Architecture},
  url = {https://www.morganclaypool.com/doi/abs/10.2200/S00874ED3V01Y201809CAC046},
}

@inproceedings{boutros2020_Peak,
  title = {Beyond {{Peak Performance}}: {{Comparing}} the {{Real Performance}} of {{AI-Optimized FPGAs}} and {{GPUs}}},
  author = {Boutros, Andrew and Nurvitadhi, Eriko and Ma, Rui and Gribok, Sergey and Zhao, Zhipeng and Hoe, James C. and Betz, Vaughn and Langhammer, Martin},
  date = {2020},
  eventtitle = {{{International Conference}} on {{Field-Programmable Technology}} ({{FPT}})}
}

@inproceedings{byma2014_FPGAs,
  title = {{{FPGAs}} in the {{Cloud}}: {{Booting Virtualized Hardware Accelerators}} with {{OpenStack}}},
  author = {Byma, Stuart and Steffan, J. Gregory and Bannazadeh, Hadi and Garcia, Alberto Leon and Chow, Paul},
  date = {2014},
  eventtitle = {International {{Symposium}} on {{Field-Programmable Custom Computing Machines}} ({{FCCM}})}
}

@inproceedings{caulfield2016_Cloudscale,
  title = {A {{Cloud-scale Acceleration Architecture}}},
  author = {Caulfield, Adrian and Chung, Eric and Putnam, Andrew and Angepat, Hari and Fowers, Jeremy and Haselman, Michael and Heil, Stephen and Humphrey, Matt and Kaur, Puneet and Kim, Joo-Young and Lo, Daniel and Massengill, Todd and Ovtcharov, Kalin and Papamichael, Michael and Woods, Lisa and Lanka, Sitaram and Chiou, Derek and Burger, Doug},
  date = {2016},
  eventtitle = {International {{Symposium}} on {{Microarchitecture}} ({{MICRO}})}
}

@inproceedings{chen2012_Using,
  title = {Using {{OpenCL}} to {{Evaluate}} the {{Efficiency}} of {{CPUs}}, {{GPUs}}, and {{FPGAs}} for {{Information Filtering}}},
  author = {Chen, Doris and Singh, Deshanand},
  date = {2012},
  url = {http://ieeexplore.ieee.org/document/6339171/},
  eventtitle = {International {{Conference}} on {{Field Programmable Logic}} and {{Applications}} ({{FPL}})}
}

@inproceedings{chen2013_Fractal,
  title = {Fractal {{Video Compression}} in {{OpenCL}}: {{An Evaluation}} of {{CPUs}}, {{GPUs}}, and {{FPGAs}} as {{Acceleration Platforms}}},
  author = {Chen, Doris and Singh, Deshanand},
  date = {2013},
  eventtitle = {Asia and {{South Pacific Design Automation Conference}} ({{ASP-DAC}})}
}

@inproceedings{chen2018_TVM,
  title = {{{TVM}}: {{An Automated End-to-End Optimizing Compiler}} for {{Deep Learning}}},
  author = {Chen, Tianqi and Moreau, Thierry and Jiang, Ziheng and Zheng, Lianmin and Yan, Eddie and Shen, Haichen and Cowan, Meghan and Wang, Leyuan and Hu, Yuwei and Ceze, Luis and Guestrin, Carlos and Krishnamurthy, Arvind},
  date = {2018},
  url = {https://www.usenix.org/conference/osdi18/presentation/chen},
  eventtitle = {{{USENIX Symposium on Operating Systems Design}} and {{Implementation}} ({{OSDI}})}
}

@article{chung2018_Serving,
  title = {Serving {{DNNs}} in {{Real Time}} at {{Datacenter Scale}} with {{Project Brainwave}}},
  author = {Chung, Eric and Fowers, Jeremy and Ovtcharov, Kalin and Papamichael, Michael and Caulfield, Adrian and Massengill, Todd and Liu, Ming and Lo, Daniel and Alkalay, Shlomi and Haselman, Michael and Abeydeera, Maleen and Adams, Logan and Angepat, Hari and Boehn, Christian and Chiou, Derek and Firestein, Oren and Forin, Alessandro and Gatlin, Kang Su and Ghandi, Mahdi and Heil, Stephen and Holohan, Kyle and El Husseini, Ahmad and Juhasz, Tamas and Kagi, Kara and Kovvuri, Ratna K. and Lanka, Sitaram and van Megen, Friedel and Mukhortov, Dima and Patel, Prerak and Perez, Brandon and Rapsang, Amanda and Reinhardt, Steven and Rouhani, Bita and Sapek, Adam and Seera, Raja and Shekar, Sangeetha and Sridharan, Balaji and Weisz, Gabriel and Woods, Lisa and Yi Xiao, Phillip and Zhang, Dan and Zhao, Ritchie and Burger, Doug},
  date = {2018},
  journaltitle = {IEEE Micro},
}

@online{cloudflare_Eliminating,
  title = {Eliminating {{Cold Starts}} with {{Cloudflare Workers}}},
  author = {Cloudflare},
  url = {https://blog.cloudflare.com/eliminating-cold-starts-with-cloudflare-workers/},
  urldate = {2021-05-16},
}

@online{cloudflare_Workers,
  title = {Workers},
  author = {Cloudflare},
  url = {https://workers.cloudflare.com/},
  urldate = {2021-05-14}
}

@online{CNCF_serverless,
  title = {{{CNCF Serverless Whitepaper}} v1.0},
  author = {CNCF Serverless Working Group},
  url = {https://github.com/cncf/wg-serverless/blob/master/whitepapers/serverless-overview/README.md},
  urldate = {2020-12-23}
}

@inproceedings{du2020_Catalyzer,
  title = {Catalyzer: {{Sub-millisecond Startup}} for {{Serverless Computing}} with {{Initialization-less Booting}}},
  author = {Du, Dong and Yu, Tianyi and Xia, Yubin and Zang, Binyu and Yan, Guanglu and Qin, Chenggang and Wu, Qixuan and Chen, Haibo},
  date = {2020},
  url = {http://doi.org/10.1145/3373376.3378512},
  eventtitle = {International {{Conference}} on {{Architectural Support}} for {{Programming Languages}} and {{Operating Systems}} ({{ASPLOS}})}
}

@inproceedings{du2022_Serverless,
  title = {{Serverless Computing on Heterogeneous Computers}},
  booktitle = {{{International Conference}} on {{Architectural Support}} for {{Programming Languages}} and {{Operating Systems} (ASPLOS)}},
  author = {Du, Dong and Liu, Qingyuan and Jiang, Xueqiang and Xia, Yubin and Zang, Binyu and Chen, Haibo},
  date = {2022},
  url = {http://doi.org/10.1145/3503222.3507732},
}

@inproceedings{eskandari2019_Modular,
  ids = {galapagos},
  title = {A {{Modular Heterogeneous Stack}} for {{Deploying FPGAs}} and {{CPUs}} in the {{Data Center}}},
  author = {Eskandari, Nariman and Tarafdar, Naif and Ly-Ma, Daniel and Chow, Paul},
  date = {2019},
  url = {https://dl.acm.org/doi/10.1145/3289602.3293909},
  eventtitle = {International {{Symposium}} on {{Field-Programmable Gate Arrays}} ({{FPGA}})}
}

@inproceedings{fowers2018_Configurable,
  title = {A {{Configurable Cloud-scale DNN Processor}} for {{Real-time AI}}},
  author = {Fowers, Jeremy and Ovtcharov, Kalin and Papamichael, Michael and Massengill, Todd and Liu, Ming and Lo, Daniel and Alkalay, Shlomi and Haselman, Michael and Adams, Logan and Ghandi, Mahdi and Heil, Stephen and Patel, Prerak and Sapek, Adam and Weisz, Gabriel and Woods, Lisa and Lanka, Sitaram and Reinhardt, Steven and Caulfield, Adrian and Chung, Eric and Burger, Doug},
  date = {2018},
  url = {http://doi.org/10.1109/ISCA.2018.00012},
  eventtitle = {International {{Symposium}} on {{Computer Architecture}} ({{ISCA}})}
}

@inproceedings{fuerst2021_FaasCache,
  title = {{{FaasCache}}: {Keeping Serverless Computing Alive with Greedy-Dual Caching}},
  booktitle = {{{International Conference}} on {{Architectural Support}} for {{Programming Languages}} and {{Operating Systems (ASPLOS)}}},
  author = {Fuerst, Alexander and Sharma, Prateek},
  date = {2021},
  url = {http://doi.org/10.1145/3445814.3446757},
}

@article{gandhi2012_AutoScale,
  title = {{{AutoScale}}: {{Dynamic}}, {{Robust Capacity Management}} for {{Multi-tier Data Centers}}},
  author = {Gandhi, Anshul and Harchol-Balter, Mor and Raghunathan, Ram and Kozuch, Michael},
  date = {2012},
  journaltitle = {ACM Transactions on Computer Systems (TOCS)},
  shortjournal = {ACM Trans. Comput. Syst.},
  url = {https://dl.acm.org/doi/10.1145/2382553.2382556},
}

@inproceedings{gujarati2017_Swayam,
  title = {Swayam: {{Distributed Autoscaling}} to {{Meet SLAs}} of {{Machine Learning Inference Services}} with {{Resource Efficiency}}},
  author = {Gujarati, Arpan and Elnikety, Sameh and He, Yuxiong and McKinley, Kathryn and Brandenburg, Björn},
  date = {2017},
  url = {https://doi.org/10.1145/3135974.3135993},
  eventtitle = {{{International Middleware Conference}} ({{Middleware}})}
}

@inproceedings{gujarati2020_Serving,
  title = {Serving {{DNNs}} like {{Clockwork}}: {{Performance Predictability}} from the {{Bottom Up}}},
  author = {Gujarati, Arpan and Karimi, Reza and Alzayat, Safya and Hao, Wei and Kaufmann, Antoine and Vigfusson, Ymir and Mace, Jonathan},
  date = {2020},
  url = {https://www.usenix.org/conference/osdi20/presentation/gujarati},
  eventtitle = {{{USENIX Symposium on Operating Systems Design}} and {{Implementation}} ({{OSDI}})}
}

@online{hoover2021_1stCLaaS,
  title = {1st-{{CLaaS}}},
  author = {Hoover, Steve},
  url = {https://github.com/stevehoover/1st-CLaaS},
  urldate = {2021-05-14}
}

@inproceedings{hosseinabady2014_Runtime,
  title = {{Run-Time Power Gating in Hybrid {{ARM-FPGA}} Devices}},
  author = {Hosseinabady, Mohammad and Nunez-Yanez, Jose Luis},
  date = {2014},
  url = {http://ieeexplore.ieee.org/document/6927503/},
  urldate = {2021-07-29},
  eventtitle = {{{International Conference}} on {{Field Programmable Logic}} and {{Applications}} ({{FPL}})},
}

@inproceedings{huang2016_Programming,
  title = {Programming and {{Runtime Support}} to {{Blaze FPGA Accelerator Deployment}} at {{Datacenter Scale}}},
  author = {Huang, Muhuan and Wu, Di and Yu, Cody Hao and Fang, Zhenman and Interlandi, Matteo and Condie, Tyson and Cong, Jason},
  date = {2016},
  eventtitle = {{{ACM Symposium on Cloud Computing}} ({{SoCC}})},
}

@online{inaccel_FPGA,
  title = {{{FPGA Manager}}},
  author = {InAccel},
  url = {https://inaccel.com/fpga-manager/},
  urldate = {2021-05-13}
}

@online{intel_OneAPI,
  title = {{{oneAPI}}: {{A New Era}} of {{Heterogeneous Computing}}},
  author = {Intel},
  url = {https://www.intel.com/content/www/us/en/developer/tools/oneapi/overview.html},
  urldate = {2022-03-06},
}

@online{intel_Stratix,
  title = {{{Stratix}}® {{V Device Datasheet}}},
  author = {Intel},
  url = {https://www.intel.com/content/dam/www/programmable/us/en/pdfs/literature/hb/stratix-v/stx5_53001.pdf},
  urldate = {2021-05-15}
}

@online{intel_Stratixa,
  title = {{{Stratix}}® 10 {{Configuration User Guide}}},
  author = {Intel},
  url = {https://www.intel.com/content/dam/www/programmable/us/en/pdfs/literature/hb/stratix-10/ug-s10-config.pdf},
  urldate = {2021-05-15}
}

@article{istvan2017_Caribou,
  title = {Caribou: {{Intelligent Distributed Storage}}},
  author = {István, Zsolt and Sidler, David and Alonso, Gustavo},
  date = {2017},
  journaltitle = {Proceedings of the VLDB Endowment},
}

@inproceedings{jia2021_Nightcore,
  title = {Nightcore: {{Efficient}} and {{Scalable Serverless Computing}} for {{Latency-Sensitive}}, {{Interactive Microservices}}},
  author = {Jia, Zhipeng and Witchel, Emmett},
  date = {2021},
  eventtitle = {International {{Conference}} on {{Architectural Support}} for {{Programming Languages}} and {{Operating Systems}} ({{ASPLOS}})},
}

@report{jonas2019_Cloud,
  title = {Cloud {{Programming Simplified}}: {{A Berkeley View}} on {{Serverless Computing}}},
  author = {Jonas, Eric and Schleier-Smith, Johann and Sreekanti, Vikram and Tsai, Chia-Che and Khandelwal, Anurag and Pu, Qifan and Shankar, Vaishaal and Menezes Carreira, Joao and Krauth, Karl and Yadwadkar, Neeraja and Gonzalez, Joseph and Popa, Raluca Ada and Stoica, Ion and Patterson, David},
  date = {2019},
  number = {UCB/EECS-2019-3},
  institution = {{EECS Department, University of California, Berkeley}},
}

@inproceedings{kannan2019_GrandSLAm,
  title = {{{GrandSLAm}}: {{Guaranteeing SLAs}} for {{Jobs}} in {{Microservices Execution Frameworks}}},
  author = {Kannan, Ram Srivatsa and Subramanian, Lavanya and Raju, Ashwin and Ahn, Jeongseob and Mars, Jason and Tang, Lingjia},
  date = {2019},
  url = {http://doi.org/10.1145/3302424.3303958},
  eventtitle = {European {{Conference}} on {{Computer Systems}} ({{EuroSys}})}
}

@inproceedings{khawaja2018_Sharing,
  title = {Sharing, {{Protection}}, and {{Compatibility}} for {{Reconfigurable Fabric}} with {{AmorphOS}}},
  booktitle = {{{USENIX Symposium on Operating Systems Design}} and {{Implementation}} ({{OSDI}})},
  author = {Khawaja, Ahmed and Landgraf, Joshua and Prakash, Rohith and Wei, Michael and Schkufza, Eric and Rossbach, Christopher},
  date = {2018}
}

@inproceedings{korolija2020_OS,
  title = {Do {{OS Abstractions Make Sense}} on {{FPGAs}}?},
  author = {Korolija, Dario and Roscoe, Timothy and Alonso, Gustavo},
  date = {2020},
  url = {https://www.usenix.org/conference/osdi20/presentation/roscoe},
  eventtitle = {{{USENIX Symposium on Operating Systems Design}} and {{Implementation}} ({{OSDI}})}
}

@article{lavasani2014_FPGAbased,
  title = {An {{FPGA-based In-line Accelerator}} for {{Memcached}}},
  author = {Lavasani, Maysam and Angepat, Hari and Chiou, Derek},
  date = {2014},
  journaltitle = {IEEE Computer Architecture Letters (CAL)},
}

@inproceedings{le2020_AlloX,
  title = {{{AlloX}}: {Compute Allocation in Hybrid Clusters}},
  author = {Le, Tan N. and Sun, Xiao and Chowdhury, Mosharaf and Liu, Zhenhua},
  date = {2020},
  url = {https://dl.acm.org/doi/10.1145/3342195.3387547},
  eventtitle = {{{European Conference}} on {{Computer Systems (EuroSys)}}},
}

@article{leland1994_Selfsimilar,
  title = {{On the Self-Similar Nature of {{Ethernet}} Traffic (Extended Version)}},
  author = {Leland, Will E. and Taqqu, Murad S. and Willinger, Walter and Wilson, Daniel V.},
  date = {1994},
  journaltitle = {IEEE/ACM Transactions on Networking (TON)},
}

@incollection{luo2021_Characterizing,
  title = {Characterizing {{Microservice Dependency}} and {{Performance}}: {{Alibaba Trace Analysis}}},
  booktitle = {{{ACM Symposium}} on {{Cloud Computing (SoCC)}}},
  author = {Luo, Shutian and Xu, Huanle and Lu, Chengzhi and Ye, Kejiang and Xu, Guoyao and Zhang, Liping and Ding, Yu and He, Jian and Xu, Chengzhong},
  date = {2021},
  url = {http://doi.org/10.1145/3472883.3487003},
}

@inproceedings{ma2020_Hypervisor,
  title = {A {{Hypervisor}} for {{Shared-Memory FPGA Platforms}}},
  author = {Ma, Jiacheng and Zuo, Gefei and Loughlin, Kevin and Cheng, Xiaohe and Liu, Yanqiang and Eneyew, Abel Mulugeta and Qi, Zhengwei and Kasikci, Baris},
  date = {2020},
  url = {http://doi.org/10.1145/3373376.3378482},
  eventtitle = {International {{Conference}} on {{Architectural Support}} for {{Programming Languages}} and {{Operating Systems}} ({{ASPLOS}})}
}

@inproceedings{manco2017_My,
  title = {My {{VM}} Is {{Lighter}} (and {{Safer}}) than Your {{Container}}},
  author = {Manco, Filipe and Lupu, Costin and Schmidt, Florian and Mendes, Jose and Kuenzer, Simon and Sati, Sumit and Yasukata, Kenichi and Raiciu, Costin and Huici, Felipe},
  date = {2017},
  url = {https://dl.acm.org/doi/10.1145/3132747.3132763},
  eventtitle = {{{Symposium on Operating Systems Principles}} ({{SOSP}})}
}

@online{martin_Microservices,
  title = {Microservices},
  author = {Martin Fowler},
  url = {https://martinfowler.com/articles/microservices.html},
  urldate = {2021-05-16}
}

@online{microsoftazure_Azure,
  title = {Azure {{Functions Premium Plan}}: {{Pre-warmed Instances}}},
  author = {Microsoft Azure},
  url = {https://docs.microsoft.com/en-us/azure/azure-functions/functions-premium-plan},
  urldate = {2021-05-16}
}

@online{amazon_ecs_capacityscaling,
  title = {Amazon {ECS} {Best Practices Guide for Capacity and Scaling}},
  author = {Amazon Web Services},
  url = {https://docs.aws.amazon.com/AmazonECS/latest/bestpracticesguide/capacity-availability.html},
  urldate = {2021-05-16}
}

@inproceedings{mishra2016_REoN,
  title = {{{{REoN}}: {{A}} Protocol for Reliable Software-Defined {{FPGA}} Partial Reconfiguration over Network}},
  author = {Mishra, Vaibhawa and Chen, Qianqiao and Zervas, Georgios},
  date = {2016},
  url = {http://ieeexplore.ieee.org/document/7857184/},
  urldate = {2021-04-01},
  eventtitle = {{{International Conference}} on {{ReConFigurable Computing}} and {{FPGAs}} ({{ReConFig}})}
}

@inproceedings{mogul2019_Nines,
  ids = {nines_not_enough},
  title = {Nines Are {{Not Enough}}: {{Meaningful Metrics}} for {{Clouds}}},
  author = {Mogul, Jeffrey and Wilkes, John},
  date = {2019},
  url = {https://dl.acm.org/doi/10.1145/3317550.3321432},
  eventtitle = {Workshop on {{Hot Topics}} in {{Operating Systems}} ({{HotOS}})}
}

@inproceedings{nurvitadhi2016_Accelerating,
  title = {Accelerating {{Recurrent Neural Networks}} in {{Analytics Servers}}: {{Comparison}} of {{FPGA}}, {{CPU}}, {{GPU}}, and {{ASIC}}},
  author = {Nurvitadhi, Eriko and {Jaewoong Sim} and Sheffield, David and Mishra, Asit and Krishnan, Srivatsan and Marr, Debbie},
  date = {2016},
  url = {http://ieeexplore.ieee.org/document/7577314/},
  urldate = {2021-03-20},
  eventtitle = {International {{Conference}} on {{Field Programmable Logic}} and {{Applications}} ({{FPL}})}
}

@inproceedings{nurvitadhi2017_Can,
  title = {Can {{FPGAs Beat GPUs}} in {{Accelerating Next-Generation Deep Neural Networks}}?},
  booktitle = {{{ International Symposium}} on {{Field-Programmable Gate Arrays (FPGA)}}},
  author = {Nurvitadhi, Eriko and Venkatesh, Ganesh and Sim, Jaewoong and Marr, Debbie and Huang, Randy and Ong Gee Hock, Jason and Liew, Yeong Tat and Srivatsan, Krishnan and Moss, Duncan and Subhaschandra, Suchit and Boudoukh, Guy},
  date = {2017},
  url = {http://doi.org/10.1145/3020078.3021740},
  urldate = {2021-05-19}
}

@online{nvidia_Triton,
  title = {Triton {{Inference Server}}},
  author = {NVIDIA},
  url = {https://developer.nvidia.com/nvidia-triton-inference-server},
  urldate = {2021-05-18}
}

@inproceedings{ojika2019_FaaM,
  ids = {ojika2019_FaaMa},
  title = {{{FaaM}}: {{FPGA-as-a-Microservice}} - {{A Case Study}} for {{Data Compression}}},
  author = {Ojika, David and Gordon-Ross, Ann and Lam, Herman and Patel, Bhavesh},
  date = {2019},
  eventtitle = {{{EPJ Web}} of {{Conferences}}}
}

@online{openfaas_OpenFaaS,
  title = {{{Serverless Functions Made Simple}}},
  author = {OpenFaaS},
  url = {https://www.openfaas.com/},
  urldate = {2021-05-14},
}

@inproceedings{ousterhout2019_Shenango,
  title = {Shenango: {{Achieving High CPU Efficiency}} for {{Latency-sensitive Datacenter Workloads}}},
  author = {Ousterhout, Amy and Fried, Joshua and Behrens, Jonathan and Belay, Adam and Balakrishnan, Hari},
  date = {2019},
  url = {https://www.usenix.org/conference/nsdi19/presentation/ousterhout},
  urldate = {2020-12-19},
  eventtitle = {{{USENIX Symposium on Networked Systems Design}} and {{Implementation}} ({{NSDI}})}
}

@inproceedings{pitchumani2015_Realistic,
  title = {{Realistic Request Arrival Generation in Storage Benchmarks}},
  author = {Pitchumani, Rekha and Frank, Shayna and Miller, Ethan L.},
  date = {2015},
  eventtitle = {{{Symposium}} on {{Mass Storage Systems}} and {{Technologies}} ({{MSST}})}
}

@inproceedings{putnam2014_Reconfigurable,
  title = {A {{Reconfigurable Fabric}} for {{Accelerating Large-scale Datacenter Services}}},
  author = {Putnam, Andrew and Caulfield, Adrian and Chung, Eric and Chiou, Derek and Constantinides, Kypros and Demme, John and Esmaeilzadeh, Hadi and Fowers, Jeremy and Gopal, Gopi Prashanth and Gray, Jan and Haselman, Michael and Hauck, Scott and Heil, Stephen and Hormati, Amir and Kim, Joo-Young and Lanka, Sitaram and Larus, James and Peterson, Eric and Pope, Simon and Smith, Aaron and Thong, Jason and Xiao, Phillip Yi and Burger, Doug},
  date = {2014},
  eventtitle = {International {{Symposium}} on {{Computer Architecuture}} ({{ISCA}})}
}

@inproceedings{qasaimeh2019_Comparing,
  title = {Comparing {{Energy Efficiency}} of {{CPU}}, {{GPU}} and {{FPGA Implementations}} for {{Vision Kernels}}},
  author = {Qasaimeh, Murad and Denolf, Kristof and Lo, Jack and Vissers, Kees and Zambreno, Joseph and Jones, Phillip},
  date = {2019},
  eventtitle = {International {{Conference}} on {{Embedded Software}} and {{Systems}} ({{ICESS}})}
}

@inproceedings{qiao2019_FPGABased,
  title = {An {{FPGA-Based BWT Accelerator}} for {{Bzip2 Data Compression}}},
  author = {Qiao, Weikang and Fang, Zhenman and Chang, Mau-Chung Frank and Cong, Jason},
  date = {2019},
  eventtitle = {International {{Symposium}} on {{Field-Programmable Custom Computing Machines}} ({{FCCM}})}
}

@inproceedings{ringlein2019_System,
  ids = {cloudfpga_pr,ringlein2019_Systema},
  title = {System {{Architecture}} for {{Network-Attached FPGAs}} in the {{Cloud}} Using {{Partial Reconfiguration}}},
  author = {Ringlein, Burkhard and Abel, Francois and Ditter, Alexander and Weiss, Beat and Hagleitner, Christoph and Fey, Dietmar},
  date = {2019},
  eventtitle = {International {{Conference}} on {{Field Programmable Logic}} and {{Applications}} ({{FPL}})}
}

@inproceedings{ringlein2020_Programming,
  title = {Programming {{Reconfigurable Heterogeneous Computing Clusters Using MPI With Transpilation}}},
  author = {Ringlein, Burkhard and Abel, Francois and Ditter, Alexander and Weiss, Beat and Hagleitner, Christoph and Fey, Dietmar},
  date = {2020},
  eventtitle = {{{International Workshop}} on {{Heterogeneous High-Performance Reconfigurable Computing}} ({{H2RC}})},
}

@inproceedings{romero2019_INFaaS,
  title={{INFaaS: Automated Model-less Inference Serving}},
  author={Romero, Francisco and Li, Qian and Yadwadkar, Neeraja J and Kozyrakis, Christos},
  booktitle={USENIX Annual Technical Conference (ATC)},
  year={2021}
}

@inproceedings{roy2022_IceBreaker,
  title = {{{IceBreaker}}: {Warming Serverless Functions Better with Heterogeneity}},
  eventtitle = {{{International Conference}} on {{Architectural Support}} for {{Programming Languages}} and {{Operating Systems (ASPLOS)}}},
  author = {Roy, Rohan Basu and Patel, Tirthak and Tiwari, Devesh},
  date = {2022},
  url = {http://doi.org/10.1145/3503222.3507750},
  urldate = {2022-03-01}
}

@inproceedings{shahrad2020_Serverless,
  title = {Serverless in the {{Wild}}: {{Characterizing}} and {{Optimizing}} the {{Serverless Workload}} at a {{Large Cloud Provider}}},
  author = {Shahrad, Mohammad and Fonseca, Rodrigo and Goiri, Íñigo and Chaudhry, Gohar and Batum, Paul and Cooke, Jason and Laureano, Eduardo and Tresness, Colby and Russinovich, Mark and Bianchini, Ricardo},
  date = {2020},
  url = {https://www.usenix.org/conference/atc20/presentation/shahrad},
  urldate = {2020-12-15},
  eventtitle = {USENIX Annual {{Technical Conference}} ({{ATC}})}
}

@inproceedings{sharma2016_Highlevel,
  title = {From {{High-level Deep Neural Models}} to {{FPGAs}}},
  author = {Sharma, Hardik and Park, Jongse and Mahajan, Divya and Amaro, Emmanuel and Kim, Joon Kyung and Shao, Chenkai and Mishra, Asit and Esmaeilzadeh, Hadi},
  date = {2016},
  eventtitle = {International {{Symposium}} on {{Microarchitecture}} ({{MICRO}})}
}

@unpublished{singhvi2019_Archipelago,
  ids = {singhvi2019_Archipelagoa},
  title = {Archipelago: {{A Scalable Low-Latency Serverless Platform}}},
  author = {Singhvi, Arjun and Houck, Kevin and Balasubramanian, Arjun and Shaikh, Mohammed Danish and Venkataraman, Shivaram and Akella, Aditya},
  year = {2019},
  eprint = {1911.09849},
  eprinttype = {arxiv},
  primaryclass = {cs},
  url = {http://arxiv.org/abs/1911.09849},
  urldate = {2021-01-12},
  archiveprefix = {arXiv}
}

@article{skhiri2019_FPGA,
  title = {From {{FPGA}} to {{Support Cloud}} to {{Cloud}} of {{FPGA}}: {{State}} of the {{Art}}},
  author = {Skhiri, Rym and Fresse, Virginie and Jamont, Jean Paul and Suffran, Benoit and Malek, Jihene},
  date = {2019},
  journaltitle = {International Journal of Reconfigurable Computing},
  url = {https://doi.org/10.1155/2019/8085461}
}

@inproceedings{suresh2019_FnSched,
  title = {{{FnSched}}: {{An Efficient Scheduler}} for {{Serverless Functions}}},
  booktitle = {{{International Workshop}} on {{Serverless Computing}}},
  author = {Suresh, Amoghvarsha and Gandhi, Anshul},
  date = {2019},
  url = {https://doi.org/10.1145/3366623.3368136},
  urldate = {2021-05-13}
}

@inproceedings{tarafdar2017_Enabling,
  ids = {openstack_fpga_cluster},
  title = {Enabling {{Flexible Network FPGA Clusters}} in a {{Heterogeneous Cloud Data Center}}},
  author = {Tarafdar, Naif and Lin, Thomas and Fukuda, Eric and Bannazadeh, Hadi and Leon-Garcia, Alberto and Chow, Paul},
  date = {2017},
  url = {http://dl.acm.org/citation.cfm?doid=3020078.3021742},
  urldate = {2020-11-27},
  eventtitle = {International {{Symposium}} on {{Field-Programmable Gate Arrays}} ({{FPGA}})}
}

@incollection{tarafdar2019_Building,
  title = {Building the {{Infrastructure}} for {{Deploying FPGAs}} in the {{Cloud}}},
  booktitle = {Hardware {{Accelerators}} in {{Data Centers}}},
  author = {Tarafdar, Naif and Lin, Thomas and Ly-Ma, Daniel and Rozhko, Daniel and Leon-Garcia, Alberto and Chow, Paul},
  date = {2019},
}

@inproceedings{tarafdar2019_LibGalapagos,
  title = {{{libGalapagos}}: {{A Software Environment}} for {{Prototyping}} and {{Creating Heterogeneous FPGA}} and {{CPU Applications}}},
  author = {Tarafdar, Naif and Chow, Paul},
  date = {2019},
  eventtitle = {{{International Workshop}} on {{FPGAs}} for {{Software Programmers (FSP Workshop)}}}
}

@inproceedings{tariq2020_Sequoia,
  title = {Sequoia: {{Enabling Quality-of-Service}} in {{Serverless Computing}}},
  author = {Tariq, Ali and Pahl, Austin and Nimmagadda, Sharat and Rozner, Eric and Lanka, Siddharth},
  date = {2020},
  url = {http://doi.org/10.1145/3419111.3421306},
  urldate = {2021-01-05},
  eventtitle = {{{ACM Symposium on Cloud Computing}} ({{SoCC}})}
}

@inproceedings{tine_SingleSource,
  title = {Single-{{Source Hardware-Software Codesign}}},
  author = {Tine, Blaise-Pascal and Yalamanchili, Sudhakar and Kim, Hyesoon},
  eventtitle = {{{Workshop on Languages, Tools, and Techniques for Accelerator Design (LATTE)}}},
  date = {2021}
}

@inproceedings{tootaghaj2015_Evaluating,
  title = {Evaluating the {{Combined Impact}} of {{Node Architecture}} and {{Cloud Workload Characteristics}} on {{Network Traffic}} and {{Performance}}/{{Cost}}},
  author = {Tootaghaj, Diman Zad and Farhat, Farshid and Arjomand, Mohammad and Faraboschi, Paolo and Kandemir, Mahmut Taylan and Sivasubramaniam, Anand and Das, Chita R.},
  date = {2015},
  eventtitle = {{{IEEE International Symposium}} on {{Workload Characterization (IISWC)}}}
}

@inproceedings{wang2002_Data,
  title = {{Data Mining Meets Performance Evaluation: Fast Algorithms for Modeling Bursty Traffic}},
  author = {Wang, Mengzhi and Madhyastha, Tara and Chan, Ngai Hang and Papadimitriou, Spiros and Faloutsos, Christos},
  date = {2002},
  eventtitle = {{{International Conference}} on {{Data Engineering (ICDE)}}},
}

@inproceedings{wang2018_Peeking,
  title = {Peeking {{Behind}} the {{Curtains}} of {{Serverless Platforms}}},
  author = {Wang, Liang and Li, Mengyuan and Zhang, Yinqian and Ristenpart, Thomas and Swift, Michael},
  date = {2018},
  url = {https://www.usenix.org/conference/atc18/presentation/wang-liang},
  urldate = {2020-12-20},
  eventtitle = {USENIX Annual {{Technical Conference}} ({{ATC}})}
}

@inproceedings{weerasinghe2016_Disaggregated,
  title = {Disaggregated {{FPGAs}}: {{Network Performance Comparison Against Bare-Metal Servers}}, {{Virtual Machines}} and {{Linux Containers}}},
  author = {Weerasinghe, Jagath and Abel, Francois and Hagleitner, Christoph and Herkersdorf, Andreas},
  date = {2016},
  eventtitle = {International {{Conference}} on {{Cloud Computing Technology}} and {{Science}} ({{CloudCom}})}
}

@inproceedings{weerasinghe2016_Networkattached,
  title = {{Network-Attached {{FPGAs}}} for {{Data Center Applications}}},
  author = {Weerasinghe, Jagath and Polig, Raphael and Abel, Francois and Hagleitner, Christoph},
  date = {2016},
  eventtitle = {International {{Conference}} on {{Field-Programmable Technology}} ({{FPT}})}
}

@inproceedings{yang2022_INFless,
  title = {{{INFless}}: {A Native Serverless System for Low-Latency, High-Throughput Inference}},
  eventtitle = {{{International Conference}} on {{Architectural Support}} for {{Programming Languages}} and {{Operating Systems (ASPLOS)}}},
  author = {Yang, Yanan and Zhao, Laiping and Li, Yiming and Zhang, Huanyu and Li, Jie and Zhao, Mingyang and Chen, Xingzhen and Li, Keqiu},
  date = {2022},
  url = {http://doi.org/10.1145/3503222.3507709},
  urldate = {2022-03-03}
}

@article{yin2015_BURSE,
  title = {{{BURSE}}: {{A Bursty}} and {{Self-Similar Workload Generator}} for {{Cloud Computing}}},
  author = {Yin, Jianwei and Lu, Xingjian and Zhao, Xinkui and Chen, Hanwei and Liu, Xue},
  date = {2015},
  journaltitle = {IEEE Transactions on Parallel and Distributed Systems (TPDS)},
}

@inproceedings{zha2020_Virtualizing,
  title = {Virtualizing {{FPGAs}} in the {{Cloud}}},
  author = {Zha, Yue and Li, Jing},
  date = {2020},
  url = {https://dl.acm.org/doi/10.1145/3373376.3378491},
  urldate = {2020-11-27},
  eventtitle = {International {{Conference}} on {{Architectural Support}} for {{Programming Languages}} and {{Operating Systems}} ({{ASPLOS}})}
}

@inproceedings{zhang2019_MArk,
  title = {{{MArk}}: {{Exploiting Cloud Services}} for {{Cost-Effective}}, {{SLO-Aware Machine Learning Inference Serving}}},
  author = {Zhang, Chengliang and Yu, Minchen and Wang, Wei and Yan, Feng},
  date = {2019},
  url = {https://www.usenix.org/conference/atc19/presentation/zhang-chengliang},
  urldate = {2020-12-19},
  eventtitle = {USENIX Annual {{Technical Conference}} ({{ATC}})}
}

@inproceedings{zhang2022_HeteroGen,
  title = {{{HeteroGen}}: {Transpiling {{C}} to Heterogeneous {{HLS}} Code with Automated Test Generation and Program Repair}},
  eventtitle = {{{International Conference}} on {{Architectural Support}} for {{Programming Languages}} and {{Operating Systems} (ASPLOS)}},
  author = {Zhang, Qian and Wang, Jiyuan and Xu, Guoqing Harry and Kim, Miryung},
  date = {2022},
  url = {http://doi.org/10.1145/3503222.3507748},
}

@inproceedings{zhao2016_Universal,
  title = {{A Universal Self-Calibrating {{Dynamic Voltage}} and {{Frequency Scaling}} ({{DVFS}}) Scheme with Thermal Compensation for Energy Savings in {{FPGAs}}}},
  author = {Zhao, Shuze and Ahmed, Ibrahim and Lamoureux, Carl and Lotfi, Ashraf and Betz, Vaughn and Trescases, Olivier},
  date = {2016},
  url = {http://ieeexplore.ieee.org/document/7468125/},
  urldate = {2021-07-29},
  eventtitle = {{{IEEE Applied Power Electronics Conference}} and {{Exposition}} ({{APEC}})}
}

@inproceedings{zhao2020_Achieving,
  title = {Achieving {{100Gbps Intrusion Prevention}} on a {{Single Server}}},
  author = {Zhao, Zhipeng and Sadok, Hugo and Atre, Nirav and Hoe, James and Sekar, Vyas and Sherry, Justine},
  date = {2020},
  url = {https://www.usenix.org/conference/osdi20/presentation/zhao-zhipeng},
  eventtitle = {{{USENIX Symposium on Operating Systems Design}} and {{Implementation}} ({{OSDI}})}
}

@inproceedings{nguyen2013agile,
  title={Agile: {Elastic Distributed Resource Scaling for Infrastructure-as-a-Service}},
  author={Nguyen, Hiep and Shen, Zhiming and Gu, Xiaohui and Subbiah, Sethuraman and Wilkes, John},
  eventtitle = {International Conference on Autonomic Computing ({{ICAC}})},
  year={2013}
}

@inproceedings{chou2019mudpm,
  title={$\mu${DPM}: {Dynamic Power Management for the Microsecond Era}},
  author={Chou, Chih-Hsun and Bhuyan, Laxmi N and Wong, Daniel},
  booktitle={International Symposium on High Performance Computer Architecture (HPCA)},
  year={2019},
}

@inproceedings{kasture2015rubik,
  title={Rubik: {Fast Analytical Power Management for Latency-critical Systems}},
  author={Kasture, Harshad and Bartolini, Davide B and Beckmann, Nathan and Sanchez, Daniel},
  booktitle={International Symposium on Microarchitecture (MICRO)},
  year={2015}
}

@inproceedings{iverson1999statistical,
  title={{Statistical Prediction of Task Execution Times Through Analytic Benchmarking for Scheduling in a Heterogeneous Environment}},
  author={Iverson, Michael A. and Ozguner, Fusun and Potter, Lee C.},
  booktitle={Heterogeneous Computing Workshop (HCW)},
  year={1999},
}

@inproceedings{gandhi2011minimizing,
  title={{Minimizing Data Center {SLA} Violations and Power Consumption via Hybrid Resource Provisioning}},
  author={Gandhi, Anshul and Chen, Yuan and Gmach, Daniel and Arlitt, Martin and Marwah, Manish},
  booktitle={International Green Computing Conference and Workshops},
  year={2011},
}

@article{hochreiter1997long,
  title={{Long Short-Term Memory}},
  author={Hochreiter, Sepp and Schmidhuber, J{\"u}rgen},
  journal={Neural computation},
  year={1997},
}

@article{nguyen2022fpga,
  title={{FPGA}-based {HPC} accelerators: {An Evaluation on Performance and Energy Efficiency}},
  author={Nguyen, Tan and MacLean, Colin and Siracusa, Marco and Doerfler, Douglas and Wright, Nicholas J and Williams, Samuel},
  journal={Concurrency and Computation: Practice and Experience},
  year={2022},
}

@article{karlin1994competitive,
  title={{Competitive Randomized Algorithms for Nonuniform Problems}},
  author={Karlin, Anna R. and Manasse, Mark S. and McGeoch, Lyle A. and Owicki, Susan},
  journal={Algorithmica},
  year={1994},
}


\end{document}